\documentclass[apj]{emulateapj} 

\usepackage{color}
\usepackage{hyperref}
\hypersetup{colorlinks,citecolor=blue}
\def\wise{\textit{WISE}}
\def\WISE{\textit{WISE}}
\def\GALEX{\textit{GALEX}}
\def\Chandra{\textit{Chandra}}
\def\Spitzer{\textit{Spitzer}}
\def\AKARI{\textit{AKARI}}
\def\IRAS{\textit{IRAS}}
\def\iraf{\textsc{IRAF}}

\def\OIII{\ion{O}{3}}
\def\NII{\ion{N}{2}}
\def\FeII{\ion{Fe}{2}}
\def\Ha{H{$\alpha$}}
\def\Hb{H{$\beta$}}
\def\sersic{S\'ersic}
\def\kms{$\rm km\,s^{-1}$}
\def\w2332{W2332$-$5056}

\begin{document}

\slugcomment{{To Be Published in The Astrophysical Journal}}

\title{WISE J233237.05$-$505643.5: a Double-Peaked Broad-Lined AGN with Spiral-Shaped Radio Morphology}

\author{
Chao-Wei Tsai\altaffilmark{1,2}, 
Thomas H. Jarrett\altaffilmark{3}, 
Daniel Stern\altaffilmark{2}, 
Bjorn Emonts\altaffilmark{4,5}, 
R. Scott Barrows\altaffilmark{6}, 
Roberto J. Assef\altaffilmark{2}, 
Ray P. Norris\altaffilmark{5}, 
Peter R. M. Eisenhardt\altaffilmark{2}, 
Carol Lonsdale\altaffilmark{7}, 
Andrew W. Blain\altaffilmark{8}, 
Dominic J. Benford\altaffilmark{9}, 
Jingwen Wu\altaffilmark{2},
Brian Stalder\altaffilmark{10}, 
Christopher W. Stubbs\altaffilmark{10},
F. William High\altaffilmark{11},
K. L. Li\altaffilmark{12}, and 
Albert K. H. Kong\altaffilmark{12}
}
 
\altaffiltext{1}{Infrared Processing and Analysis Center, California Institute of Technology, Pasadena, CA 91125, USA}
\altaffiltext{2}{Jet Propulsion Laboratory, California Institute of Technology, 4800 Oak Grove Dr., Pasadena, CA 91109, USA; [email: Chao-Wei.Tsai@jpl.nasa.gov]}
\altaffiltext{3}{Astronomy Department, University of Cape Town, Private Bag X3, Rondebosch 7701, South Africa}
\altaffiltext{4}{Centro de Astrobiolog\'{i}a (INTA-CSIC), Ctra de Torrej\'{o}n a Ajalvir, km 4, 28850 Torrej\'{o}n de Ardoz, Madrid Spain}
\altaffiltext{5}{Australia Telescope National Facility, CSIRO Astronomy and Space Science, PO Box 76, Epping NSW, 1710, Australia}
\altaffiltext{6}{Arkansas Center for Space and Planetary Sciences, University of Arkansas, Fayetteville, AR 72701, USA}
\altaffiltext{7}{National Radio Astronomy Observatory, 520 Edgemont Road, Charlottesville, VA 22903, USA}
\altaffiltext{8}{Department of Physics \& Astronomy, University of Leicester, 1 University Road, Leicester, LE1 7RH, UK}
\altaffiltext{9}{NASA Goddard Space Flight Center, Greenbelt, MD 20771, USA}
\altaffiltext{10}{Harvard-Smithsonian Center for Astrophysics, 60 Garden Street, Cambridge, MA 02138, USA}
\altaffiltext{11}{Kavli Institute for Cosmological Physics, University of Chicago, 5640 South Ellis Avenue, Chicago, IL 60637, USA}
\altaffiltext{12}{Institute of Astronmy and Department of Physics, National Tsing Hua University, Hsinchu 30013, Taiwan}

\begin{abstract} 
We present radio continuum mapping, optical imaging and spectroscopy of the newly discovered double-peaked broad-lined AGN WISE J233237.05$-$505643.5 at redshift $z = 0.3447$. This source exhibits an FR-I and FR-II hybrid-morphology, characterized by bright core, jet, and Doppler-boosted lobe structures in ATCA continuum maps at 1.5, 5.6, and 9 GHz. Unlike most FR-II objects, \w2332 is hosted by a disk-like galaxy. The core has a projected 5\arcsec\ linear radio feature that is perpendicular to the curved primary jet, hinting at unusual and complex activity within the inner 25 kpc. The multi-epoch optical-near-IR photometric measurements indicate significant variability over a 3--20 year baseline from the AGN component. Gemini-South optical data shows an unusual double-peaked emission-line features: the centroids of the broad-lined components of \Ha\ and \Hb\ are blueshifted with respect to the narrow lines and host galaxy by $\sim$ 3800 \kms. We examine possible cases which involve single or double supermassive black holes in the system, and discuss required future investigations to disentangle the mystery nature of this system. 

\end{abstract}

\keywords{Galaxies: individual: WISE J233237.05$-$505643.5; Galaxies: jets; Galaxies: nuclei; Galaxies: active; Galaxies: interactions; Radio continuum: galaxies}

\section{Introduction}\label{sec:intro}

The discovery of supermassive black holes (SMBHs) with masses greater than 10 billion solar masses in two nearby galaxies \citep{2011Natur.480..215M,2012ApJ...756..179M} highlights the need to understand how SMBHs grow and influence galaxy evolution. Two commonly accepted growth processes are accretion from the interstellar medium (ISM) and SMBH mergers \citep{2012NewAR..56...93A}. The latter scenario is thought to play a more dominant role for SMBH growth in the early universe, though binary AGN seen at late stages in the merging of bulge-dominated galaxies should still be relatively common in the nearby Universe. However, despite significant efforts searching for resolved binary AGN \citep[e.g.,][]{2003ApJ...582L..15K,2006ApJ...646...49R,2010ApJ...717..209C,2010ApJ...710.1578G,2011ApJ...733..103F,2012ApJ...745...67F,2012ApJ...753...42C,2012ApJ...744....7B} and late-stage (unresolved) merging SMBHs \citep{2013arXiv.1306.4987J,2013arXiv1306.4330S}, such systems appear to be very rare. Two potential signatures of merging SMBHs are from their complicated radio morphology \citep[see][]{2011PhT....64j..32B} and unusual spectroscopic features such as velocity offsets between broad and narrow spectral line components \citep[see review by][]{2012NewAR..56...74P,2012AdAst2012E..14K}.  

SBMH binaries are the inevitable result of galaxy formation. In a major galaxy merger event, the SMBHs hosted in two large merging galaxies dynamically interact as the parent galaxies coalesce and form a final bulge. SMBH merging can be considered in three stages \citep[see][and references therein]{2005LRR.....8....8M,2008ApJS..174..455B}. In the first stage, the SMBHs dissipate their angular momentum through dynamical friction with surrounding stars, thus reducing their orbits. This process is efficient near the nuclear region of the host galaxy where the stellar density is high, but stalls at radii of $\sim$ 0.01--10 pc due to the depletion of stars. SMBHs spend the majority of their lifetime at this close separation \citep{1980Natur.287..307B,2002MNRAS.331..935Y}. However, if the system contains a gaseous accretion disk, the binary orbit could decay further via interaction between the SMBH binary and the gas \citep{2004ApJ...607..765E,2005ApJ...630..152E,2007MNRAS.379..956D,2009MNRAS.396.1640D}. This is the so-called ``intermediate stage'', lasting for a few to a few dozen Myr. In the final stage, when two SMBHs are brought close enough, the binary orbital angular momentum dissipates efficiently via gravitational radiation and the SMBHs proceed to final coalescence.

Multi-jet radio morphology is potential strong evidence for SMBH binaries or SMBH mergers. In the case of 3C 75 ($z=0.0232$), four radio jets are launched from a pair of accreting AGNs separated by 8 kpc \citep{1985ApJ...294L..85O,1985PASJ...37..655Y}. Another case, the radio galaxy 3C 66B ($z = 0.0215$), shows zig-zag patterns of compact radio jets associated with an SMBH binary, separated by $4 \times 10^{-3}$ pc in a 1.1 year orbital period \citep{2003Sci...300.1263S}. \cite{2010ApJ...724L.166I} calculated the decay time of 3C 66B orbit is about 500 yr due to gravitational radiation, although a pulsar timing experiment did not detect the expected gravitational wave from this binary to a 95\% confidence level \citep{2004ApJ...606..799J}. From their short lifetimes, SMBH binary systems with small orbital separations and short orbital periods are expected to be very rare. X-shaped radio galaxies \citep[XRGs;][]{1992ersf.meet..307L}, a subclass of Fanaroff-Riley Class II \citep[FR-II;][]{1974MNRAS.167P..31F} objects, contain two misaligned pairs of centro-symmetric radio lobes originating from an elliptical galaxy host. Usually one pair of more extended lobes with low surface brightness is oriented at a large angle with respect to a pair of less extended, high surface brightness lobes. Their radio morphology has been suggested as a sign from rapidly changing SMBH spin, possibly tracing an SMBH merging event \citep{2002MNRAS.330..609D,2002Sci...297.1310M,2003ApJ...594L.103G}. Other proposed explanations for XRGs include back-flowing material from jets deflected off ISM, outflow from over-pressured cocoons, or interactions of jets with stellar and intergalactic shells \citep{1984MNRAS.210..929L,1995ApJ...449...93W,2002A&A...394...39C,2010ApJ...710.1205H}. These models have been tested in X-ray and optical spectroscopic morphological studies \citep[e.g.][]{2009ApJ...695..156S,2009ApJS..181..548C,2010MNRAS.408.1103L,2010ApJ...710.1205H}. A recent review of observational evidence argues that none of the proposed models can explain all of the observed properties of XRGs \citep{2010arXiv1008.0789G}.

In optical spectroscopy, host galaxies with broad emission lines significantly Doppler-shifted from the corresponding narrow lines have been interpreted as the consequence of a late stage SMBH merger, or alternatively, due to a kick received by the broad-lined SMBH resulting from anisotropic gravitational emission \citep{2007PhRvL..98w1101G,2010ApJ...717..209C}. Theoretical models predict a maximum ``kick'' velocity of $\sim$ 4000 \kms\ for a merger inspiraling on a quasi-circular orbit \citep{2007ApJ...659L...5C}. In certain unusual orbit configurations such as hyperbolic encounters, the velocity displacement $\Delta V$ between narrow lines (NLs) and broad lines (BLs) can reach 10,000 \kms\ \citep{2009PhRvL.102d1101H}. Observationally, however, large BL-NL velocity offsets are not commonly seen. In the past few years, studies that mine the Sloan Digital Sky Survey (SDSS) spectroscopic archive have identified only 16 objects whose broad Balmer lines are offset from the narrow lines by $>$ 3500 km s$^{-1}$ \citep{2003ApJ...582L..15K,2009Natur.458...53B,2009ApJ...707..936S,2012arXiv1209.1635S,2011ApJ...738...20T,2012ApJS..201...23E}. In addition, two alternate models have also been proposed to explain this unusual spectral configuration in these objects: (1) a line-of-sight superposition of two unrelated AGN \citep[e.g.,][]{2009ApJ...707..936S}, and (2) an asymmetric Keplerian accretion disk \citep{1989ApJ...344..115C}. 

The accretion disk model, which requires a centrally-illuminated, geometrically thin, and optically thick accretion disk, has successfully explained the double-peaked BL emitters \citep[``double-peaked emitters'' or DPEs;][]{2007ApJ...666L..13B,2009ApJ...696.1367S,2009NewAR..53..133E}. It is expected that accretion disks should naturally become circular after a period of time \citep{1988Natur.333..523R,1990ApJ...351...38C}, and for this reason the circular, Keplerian disk model of \cite{1989ApJ...344..115C} has been used with success on many DPEs.  However, an elliptical accretion disk \citep{1995ApJ...438..610E} might arise if an instability is introduced in the disk, such as tidal disruption of a star or an orbiting body (such as a second black hole) which might allow an elliptical accretion disk to persist for some time \citep{1992MNRAS.255...92S}. The disk profiles are typically centered around Balmer emission line positions presumably because of photoionization and electron scattering in the high-density medium of the disk.  There is typically a peak on the red side and a peak on the blue side of the narrow emission lines. The specific shape of the line profile is dependent on geometrical factors such as the disk inclination, inner and outer radii, line broadening, and surface emissivity. The blue peaks can sometimes be stronger than the red peaks because of modest relativistic beaming. 

Other configurations of an accretion disk can also produce similar double-peaked BL features. For example, asymmetric BL profiles can raise from non-uniformly distributed gas such as that in long-lived eccentric accretion disks \citep{2001MNRAS.325..231O}. The warped accretion disk, produced by radiation-driven instability \citep{1996ApJ...472..582M}, can generate the double-peaked line emission profile \citep{2008MNRAS.389..213W}. In addition, a one-arm spiral excited by self-gravity in the accretion disk can also create asymmetry in the line profiles \citep{2003ApJ...598..956S}. These configurations, although all capable of producing double-peaked BL features, predict different detailed line profiles and variability, and can be tested by long-term spectroscopic monitoring \citep{2007ApJS..169..167G,2010ApJS..187..416L}. A few other complex scenarios involving two distinct orbiting SMBHs have also been proposed to explain a few individual DPEs with unusual profiles \citep{2009ApJ...704.1189T,2011NewA...16..122B}.

In this paper, we present the newly discovered double-peaked emitter WISE J233237.05$-$505643.5 (\w2332\ hereafter) at $z = 0.3447$. \w2332\ has a radio luminosity of $2.0 \times 10^{25}$ W\,Hz$^{-1}$\,str$^{-1}$ at 4.85 GHz, comparable to other FR-II galaxies \citep[][assuming a spectral index $\alpha = -0.8$ between 4.85 GHz and 178 MHz]{1974MNRAS.167P..31F}. The radio continuum morphology of this object is complex, exhibiting both FR-I and FR-II characteristics in addition to a winding primary jet with multiple structures near the core, including a linear structure perpendicular to the primary jet. The optical spectrum of \w2332\ exhibits broad emission lines (\Ha\ and \Hb) blueshifted by $\sim$ 3800 km s$^{-1}$ with respect to the corresponding narrow lines. The combination of these features makes \w2332 a prime candidate for an SMBH merger system. 

We describe the multi-band observations of \w2332 at mid-infrared (mid-IR), radio, optical, and X-ray wavelengths in  \S \ref{sec:observation}. The detailed results comprise \S \ref{sec:results_and_analyses}. In \S \ref{sec:discussion}, we interpret our results and discuss various interpretations of this unusual object. We summarize our work in  \S \ref{sec:summary}. Throughout this paper, we assume a concordance cosmology with $H_0 = 70\, {\rm km}\, {\rm s}^{-1}\, {\rm Mpc}^{-1}$, $\Omega_m = 0.3$, and $\Omega_\Lambda = 0.7$. For this cosmology, 1\arcsec\ at $z=0.3447$ subtends 4.9 kpc.

\section{Source Selection and Observations}\label{sec:observation}

\w2332\ was identified based on its unusual mid-IR colors from the NASA {\it Wide-field Infrared Survey Explorer} \citep[\wise;][]{2010AJ....140.1868W}. \WISE\ mapped the entire sky in four mid-IR bands  (3.4, 4.6, 12 and 22 \micron, or W1, W2, W3, and W4, respectively), achieving typical point-source sensitivities of 0.08, 0.11, 1 and 6 mJy  (5$ \sigma$) in the four bands, respectively. In the short-wavelength bands, \w2332\ has a relatively red [3.4]$-$[4.6] color of 0.9 mag, similar to normal AGN \citep{2012ApJ...753...30S} and occupies the same color-color space as other SMBH merger/recoiling candidates discussed in \S \ref{sec:intro}. However, the [4.6]$-$[12] color is 2.0 mag, significantly bluer than the typical AGN seen by \wise\ \citep{2011ApJ...735..112J}. It is also a strong radio emitter, having been previously detected at 4.85 GHz in the PMN Survey \citep{1994ApJS...90..173G,1994ApJS...91..111W} and at 843 MHz in the SUMSS survey \citep{2003MNRAS.342.1117M}. We obtained higher spatial resolution radio follow-up observations using the Australian Telescope Compact Array (ATCA), and optical imaging and spectroscopy with the Magellan, SOAR and Gemini South telescopes.  Details of the observations are given below.

\w2332\ is 8.7 arcminutes SE of the background $z = 0.5707$ galaxy cluster SPT-CL J2331$-$5051, which was discovered by the South Pole Telescope (SPT) Sunyaev-Zel'dovich survey \citep{2010ApJ...722.1180V}. As a result, it has been observed with Magellan optical broadband imaging \citep[Inamori Magellan Areal Camera and Spectrograph, or IMACS;][]{2011PASP..123..288D}, the \Spitzer\ IRAC camera, and the \Chandra\ ACIS-I camera as part of SPT cluster follow-up programs. These observations are detailed below, and aperture photometry is presented in Table \ref{table:w2332_photo}.

\begin{deluxetable*}{clllclll}  
\tabletypesize{\scriptsize}
\tablewidth{0in}
\tablecaption{Multi-Wavelength Photometry of \w2332\label{table:w2332_photo}}
\tablehead{
\multicolumn{1}{c}{Band} &
\multicolumn{1}{c}{R.A. (J2000.0)} &
\multicolumn{1}{c}{Decl. (J2000.0)} &
\multicolumn{1}{l}{Wavelength} &
\multicolumn{1}{c}{Instrument} &
\multicolumn{1}{c}{Flux Density} &
\multicolumn{1}{c}{Epoch (UT)} &
\multicolumn{1}{c}{Notes}
}
\startdata
Hard X-ray & 23$^\textrm{h}$32$^\textrm{m}$37$^\textrm{s}$ & -50\arcdeg56\arcmin43\arcsec	& 2--10 keV 	& \textit{Chandra}/ACIS	& $53^{+8}_{-7}$ nJy\tablenotemark{a}	 & 2009 August 12, 30 \\
Soft X-ray & ... & ...																	& 0.5--2 keV 	& ... 		& $63^{+13}_{-11}$ nJy\tablenotemark{a}	 & ... \\		
FUV		& 23$^\textrm{h}$32$^\textrm{m}$36$\fs$89 	& -50\arcdeg56\arcmin42\farcs8			& 1516 \AA	& \textit{GALEX	}& 6$\pm$3 $\mu$Jy& $\rceil$ 2003 Aug. 10 \\
NUV 	& ... & ...																	& 2267 \AA	& ...		& 7$\pm$2 $\mu$Jy& $\rfloor$ -- 2005 Sep. 06\\
UKST-$B_{j}$	& 23$^\textrm{h}$32$^\textrm{m}$37$\fs$06 	& -50\arcdeg56\arcmin43\farcs2			& 4400 \AA	& SuperCOSMOS\tablenotemark{a}	& 26$\pm$6 $\mu$Jy& 1975 July 16	& \\
UKST-$R$	& ... & ...																	& 6600 \AA	& ...		& 90$\pm$20 $\mu$Jy& 1992 November 14 & \\
UKST-$I$	& ... & ...																	& 8000 \AA	& ...		& 80$\pm$20 $\mu$Jy& 1991 August 03 & \\
$g^\prime$		& 23$^\textrm{h}$32$^\textrm{m}$36$\fs$96 	& -50\arcdeg56\arcmin43\farcs44			& 4770 \AA 	& Magellan/IMACS\tablenotemark{b} &   85.1$\pm$0.5 $\mu$Jy 		& 2008 November 3--4 \\
$r^\prime$ 		& ... & ... 																	& 6231 \AA 	& ... 				& 182$\pm$1 $\mu$Jy 	&	...\\
$i^\prime$ 		& ... & ... 																	& 7625 \AA 	& ... 				& 254$\pm$1 $\mu$Jy	&	... \\
$z^\prime$ 		& ... & ... 																	& 9134 \AA 	& ... 				& 432$\pm$1 $\mu$Jy	&	...\\
$J$ 		& 23$^\textrm{h}$32$^\textrm{m}$37$\fs$06 	& -50\arcdeg56\arcmin43\farcs3			& 1.235 \micron 	& 2MASS 	& 0.37$\pm$0.06 mJy	& 1999 October 18	\\
$H$ 		& ... & ... 																	& 1.662 \micron 	& ... 		& 0.42$\pm$0.09 mJy	& ...	\\
$K_\textrm{s}$ 		& ... & ... 															& 2.159 \micron 	& ... 		& 0.79$\pm$0.09 mJy	& ...	\\
IRAC Channel 1 	& 23$^\textrm{h}$32$^\textrm{m}$37$\fs$12 	& -50\arcdeg56\arcmin44\farcs1			& 3.6 \micron 	& \textit{Spitzer}/IRAC 	& 1.311$\pm$0.002 mJy	&	2009 November 26\\
W1 		& 23$^\textrm{h}$32$^\textrm{m}$37$\fs$05 	& -50\arcdeg56\arcmin43\farcs5			& 3.4 \micron 	& \textit{WISE} 	& 1.31$\pm$0.03 mJy	&	2010 May 19--20 \\
W2 		& ... & ... 																	& 4.6 \micron 	& ... 		& 1.70$\pm$0.04 mJy	&	... \\
W3 		& ... & ... 																	& 12 \micron 	& ... 		& 2.0$\pm$0.1 mJy	&	... \\
W4 		& ... & ... 																	& 22 \micron 	& ... 		& 4$\pm$1.0 mJy	&	... \\
\AKARI-9\micron 	&    &   																	&  9 \micron  	& \textit{AKARI} 	& $<$ 0.03 Jy\tablenotemark{c}	&	--	& FWHM = 5\farcm5\\
\AKARI-18\micron 	&    &   																	&  18 \micron  	& ... 	& $<$ 0.05 Jy\tablenotemark{c}	&	--	& FWHM = 5\farcm7\\
\IRAS-60\micron 	&    &   															&  60 \micron  	& \textit{IRAS} 	& $<$  0.5 Jy	&	--	& FWHM = 1\farcm5 $\times$ 4\farcm8 \\
\IRAS-100\micron 	&    &   															&  100 \micron  & ... 	& $<$  1.5 Jy	&	--	& FWHM = 3\farcm0 $\times$ 5\farcm1 \\
150 GHz 	&    &   																	&  2 mm  	& SPT Survey\tablenotemark{d} 	& $<$ 2.6 mJy\tablenotemark{c}	&	--	& FWHM = 1\arcmin\\
9.0 GHz   & 23$^\textrm{h}$32$^\textrm{m}$37$\fs$06 	& -50\arcdeg56\arcmin43\farcs6			&  3.3 cm  	& ATCA 						& 37.2$\pm$0.5 mJy\tablenotemark{e} 	&  2010 July 18--19 & $\theta_{\rm LAS} = 2\farcm3$\\
5.6 GHz   & 23$^\textrm{h}$32$^\textrm{m}$37$\fs$05 	& -50\arcdeg56\arcmin43\farcs3			&  5.4 cm  	& ... 						& 48.6$\pm$0.3 mJy\tablenotemark{e} 	&  ... 		& $\theta_{\rm LAS} = 3\farcm9$\\
1.5 GHz   & 23$^\textrm{h}$32$^\textrm{m}$36$\fs$86 	& -50\arcdeg56\arcmin41\farcs3			&  20 cm  		& ... 						& 103$\pm$2 mJy\tablenotemark{e} 	&  2010 July 16--17 & $\theta_{\rm LAS} = 5\farcm5$\\
4.85 GHz 	& 23$^\textrm{h}$32$^\textrm{m}$37$\fs$05 	& -50\arcdeg56\arcmin43\farcs5			&  6.18 cm  	& PMN Survey\tablenotemark{f} 	& 47$\pm$9 mJy		&	--	& FWHM = 45\arcsec\\
843 MHz 	& 23$^\textrm{h}$32$^\textrm{m}$36$\fs$62 	& -50\arcdeg56\arcmin42\farcs6			& 35.6 cm		& SUMSS\tablenotemark{g} 	        & 175$\pm$5 mJy	&	--	& FWHM = 4\farcm2\\
\enddata
\tablenotetext{a}{Corresponding to the observed $f(0.5-2\, {\rm keV}) =2.32^{+0.34}_{-0.32}\, \times 10^{-13}\, {\rm erg}\, {\rm cm}^{-2}\,{\rm s}^{-1}$, and $f(2-10\, {\rm keV})= 1.01^{+0.20}_{-0.17}\, \times 10^{-12}\, {\rm erg}\, {\rm cm}^{-2}\,{\rm s}^{-1}$.}
\tablenotetext{b}{\cite{2001MNRAS.326.1295H}}
\tablenotetext{c}{$3 \sigma$ limit.}
\tablenotetext{d}{Joaquin Vieira, private communication.}
\tablenotetext{e}{Integrated flux within 0\farcm5 radius.}
\tablenotetext{f}{\cite{1994ApJS...90..173G,1994ApJS...91..111W}}
\tablenotetext{g}{\cite{2003MNRAS.342.1117M}}
\end{deluxetable*}  

\subsection{Radio Observations}

The radio continuum observations were carried out at ATCA using the Compact Array Broadband Backend (CABB) with 2$\times$2 GHz bandwidth. \w2332\ was mapped at 1.5 GHz in the 6C configuration on UT 2010 July 16-17. The 5.6 GHz and 9.0 GHz data were obtained simultaneously on UT 2010 July 18-19 in the 6C configuration (maximum baseline $\sim$ 6.0 km), and on UT 2010 August 10 in the hybrid H168 configuration (with 5 antennas closely positioned within 200\,m of each other, and the 6th antenna at 4.5\,km distance). Observations were taken in snap-shot mode over 12h (6C) and 6h (H168) periods. A total of 25 minutes integration was obtained at 1.5 GHz, and 166 minutes at 9.0 GHz and 5.6 GHz. PKS 1934-63 was used as the flux calibrator while PKS 2311-452 (for 5.6 GHz and 9.0 GHz) and PKS 2333-528 (all three bands) were used for phase calibration. The largest angular scales ($\theta_{\rm LAS}$) visible under these configurations are 5\farcm5 at 1.5 GHz, 3\farcm9 at 5.6 GHz, and 2\farcm3 at 9.0 GHz. 

\begin{figure*}[ht!]
\figurenum{1}
\epsscale{0.95}
\begin{center}
\plotone{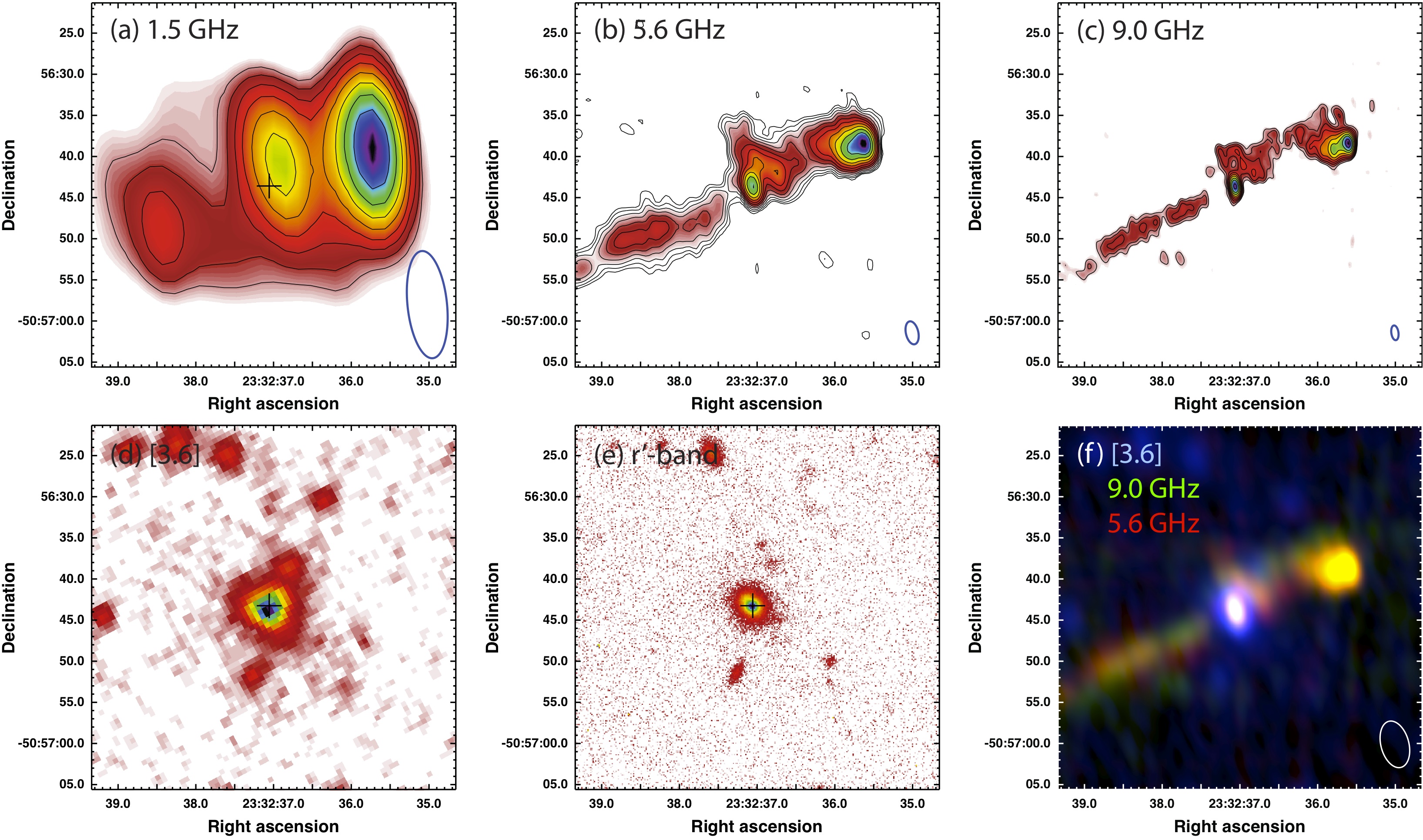}
\end{center}
\caption{
\w2332\ imaging at radio, near-IR, and optical wavelengths. Images are in (\textit{a}) 1.5 GHz continuum (beam size: $13\farcs1 \times 4\farcs8$, PA = 6\degr); (\textit{b}) 5.6 GHz continuum ($2\farcs85 \times 1\farcs51$, PA = 13\degr) ; (\textit{c}) 9.0 GHz continuum ($1\farcs80 \times 0\farcs87$, PA = 8\degr); (\textit{d}) \Spitzer\ IRAC 3.6 \micron\ (FWHM $\sim$ $1\farcs7$), and (\textit{e}) $r^{\prime}$-band (seeing = $0\farcs65$). In the radio maps (\textit{a-c}), the colors represent the flux density from peak in each map to $3\sigma$ level, and the beam sizes of maps are illustrated with blue ellipses in the lower-right corner of each panel. The contour levels are at $2^{0.5 n} \times 5 \sigma$, where $n$ is non-negative integers, and $1\sigma$ are 0.47 mJy/beam  at 1.5 GHz, 0.026 mJy/beam at 5.6 GHz, and 0.034 mJy/beam at 9.0 GHz. In the 3.6 \micron\ (\textit{d}) and $r^{\prime}$-band images (\textit{e}), the colors show the flux density peak from background levels on a logarithmic scale. The location of the 9 GHz radio core is marked with a ``+'' in (a), (d), and (e). The final panel (f) shows a composite mid-IR and radio view of \w2332\ (blue: \Spitzer\ IRAC 3.6 \micron; green: 9.0 GHz; red: 5.6 GHz) in the matched resolution ($2\farcs9 \times 1\farcs7$, PA = 15\degr). 
}\label{fig:radio_map}\label{fig:optical_image}\label{fig:irac_image}
\end{figure*}

The data reduction was done using {\sc miriad}. The final robust-weighted images are shown in Figure \ref{fig:radio_map} (a)--(c). The primary beam at the highest frequency, 9.0 GHz, is $> 6\arcmin$, significantly larger than the 1\arcmin\ field of interest for \w2332, and hence we did not apply a primary beam correction. The rms of the 1.5 GHz map is 0.62 mJy\,beam$^{-1}$ with a beam size of 13\farcs12 $\times$ 4\farcs80, PA = 6\degr. The (\textit{u,v}) data at 5.6 GHz and 9.0 GHz from the two configurations were combined, resulting in an rms noise of 29 $\mu$Jy\,beam$^{-1}$ and 37 $\mu$Jy\,beam$^{-1}$ in the final synthesized beams of 2\farcs85 $\times$ 1\farcs51, PA = 13\degr and 1\farcs80 $\times$ 0\farcs87, PA = 8\degr, respectively. The results of the integrated flux measurements using {\sc aips} are listed in Table \ref{table:w2332_photo}. Because the $\theta_{\rm LAS}$ for each band is larger than the angular extent of \w2332, we anticipate negligible or minimal missing flux in the ATCA interferometric measurements. 

\subsection{Mid-IR Observations}

Table \ref{table:w2332_photo} presents mid-IR photometry of \w2332\ from the \WISE\ All-Sky Release Atlas Catalog \citep{2012wise.rept....1C}. The observations of this field were carried out on UT 2010 May 19--20. A total of 12 single frames in each \WISE\ band are coadded. The spatial resolution of \WISE\ is  $\sim$6\arcsec\ in W1, W2, and W3, and $\sim$12\arcsec\ in W4. \w2332\ is well detected in W1$-$W3, with signal-to-noise ratio (SNR) $>$ 15, and marginally detected (4 $\sigma$) in W4. The mid-IR fluxes are listed in Table \ref{table:w2332_photo}.

In addition, \w2332 was observed in W1 and W2 during the \WISE\ post-cryogenic survey on UT 2010 November 13, 16, 17, and 18. A total of 19 frames were acquired, spread across 3 days due to the satellite maneuvering to avoid the Moon during its Sun-synchronous polar orbit. These 2nd epoch measurements allow us to test the variability of the AGN in \w2332\ at mid-IR wavelengths, as discussed in \S \ref{sec:variability}.

Finally, due its projected proximity to SPT-CL J2331$-$5051, \w2332 was also observed in [3.6] during \Spitzer\ Cycle-9 (ID: 60099, PI: Mark Brodwin) on UT 2009 November 26. The pipeline-processed data (PBCD; shown in panel (d) of Figure \ref{fig:irac_image}), retrieved from the \Spitzer\ Archive, provides another mid-IR epoch for comparison.

\subsection{Optical and Near-IR Observations}

\subsubsection{Optical and Near-IR Imaging}
Optical photometry of \w2332\ was obtained with Magellan-IMACS in the SDSS \textit{g$^\prime$r$^\prime$i$^\prime$z$^\prime$}-bands as part of SPT cluster follow-up efforts \citep{2010ApJ...722.1180V}. The observations were taken on UT 2008 November 3-4 with median seeing of $\sim$ 0\farcs8, and the 5-$\sigma$ magnitude limits in AB system are $\sim$ 24.8, 24.8, 24.4, and 23.4 for \textit{g$^\prime$}, \textit{r$^\prime$}, \textit{i$^\prime$}, and \textit{z$^\prime$}, respectively. The details of the Magellan observations and data reduction are described in \cite{2010ApJ...723.1736H}. 

To explore variability, we obtained archival photometric data of \w2332\ from several major sky surveys. The optical photometry from SuperCOSMOS Sky Survey \citep{2001MNRAS.326.1295H} in UKST $B_{j}$-, $R$-, and $I$-bands has a photometric uncertainty of $\sim 0.3$ mag. Near-IR imaging from the Two Micron All-Sky Survey \citep[2MASS;][]{2006AJ....131.1163S} was obtained on UT 1999 October 18 and provided 5 -- 9 $\sigma$ detections in the $J$-, $H$-, and $K_s$-bands. Detailed information of these sky survey observations is presented in Table \ref{table:w2332_photo}.

In order to study the morphology of \w2332, we also acquired high quality \textit{$r^\prime$}-band imaging ($\bar{\lambda} \sim$ 6165\AA) with 0\farcs65 seeing using the 4.1m Southern Astrophysical Research (SOAR) Telescope with the SOAR Optical Imager \citep[SOI;][]{2003SPIE.4841..286W} on UT 2011 November 22. The basic image processing and reductions were carried out with \iraf. This $r^{\prime}$-band image is shown in panel (e) of Figure \ref{fig:optical_image}.

\subsubsection{Optical Spectroscopy}
 
\begin{figure*}
\figurenum{2}
\begin{center}
\plotone{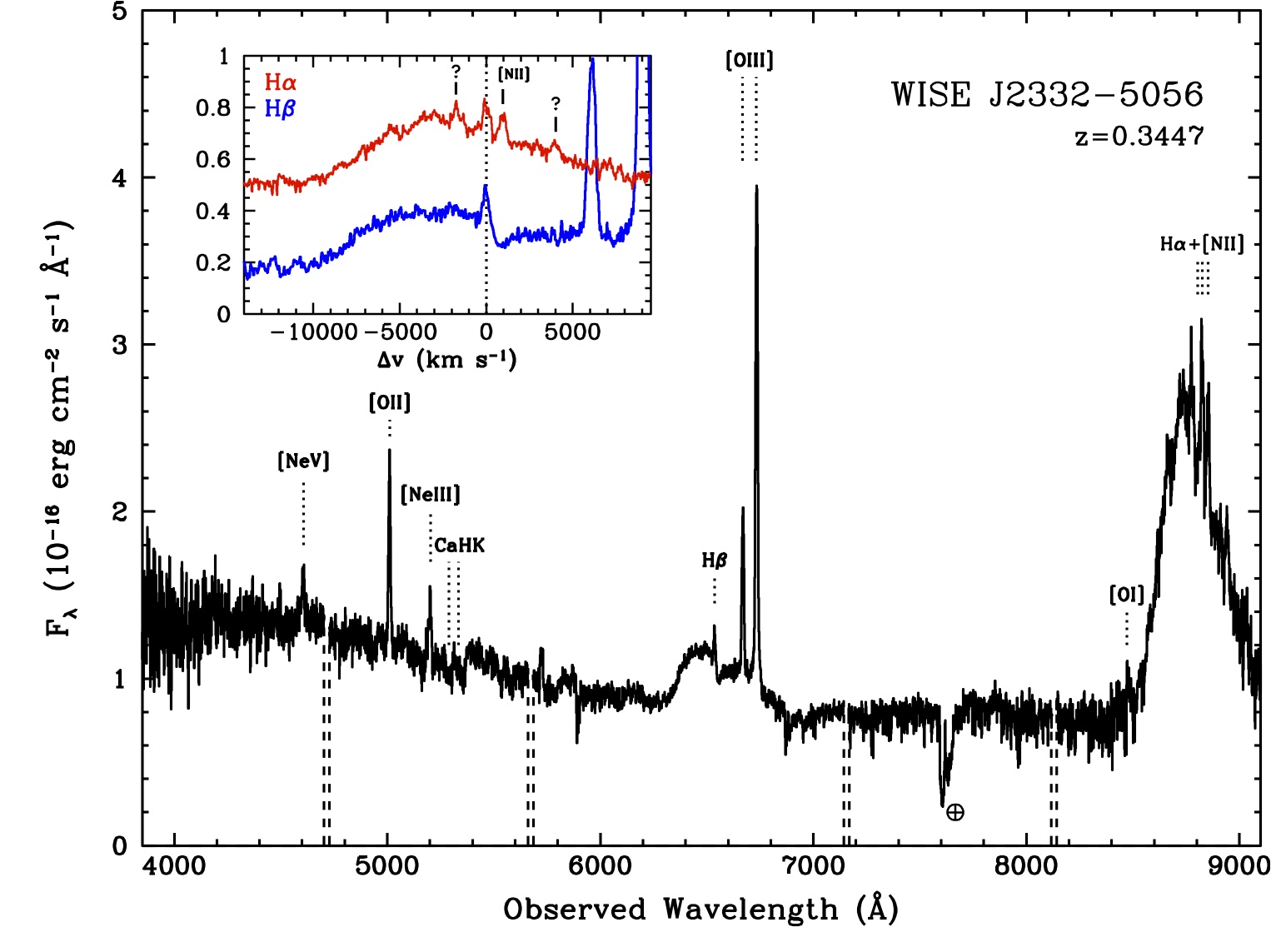}
\end{center}
\caption{Combined GMOS-S optical spectrum of \w2332 from 3800\AA\ to 9100\AA. The narrow lines are marked at redshift $z=0.3447$. The sign ``$\earth$'' labels the atmosphere absorption features. The scaled broad \Ha\ and \Hb\ lines are highlighted in the inset panel, indicating velocities relative to the narrow line redshift (shown by the dotted vertical line). The dashed lines around 4715\AA, 5675\AA, 7155\AA, and 8130\AA\ represent the CCD gaps of GMOS-S. The signs ``?'' show the unidentified narrow-band features.
}\label{fig:optical_spec}
\end{figure*}

The optical spectroscopic observations of \w2332\ were carried out with the Gemini Multi-Object Spectrograph \citep[GMOS;][]{2004PASP..116..425H} on the Gemini South telescope (Gemini-S) on UT 2011 May 11 and UT 2011 November 28. We obtained two 10-minute integrations observations with the 1\farcs5 slit and the B600/520.0 disperser in May 2011, covering 3800\AA\ to 6700\AA\ with resolving power $R \equiv \lambda/\Delta\lambda = 1700$. At $z = 0.3447$, the \Hb\ line is near the red end of the wavelength coverage. In November 2011, in order to cover \Hb\ and \Ha\ together and confirm the broad line \Hb\ feature observed earlier, we revisited \w2332\ with Gemini-S/GMOS, with the 1\farcs5 slit and the R400/705.0 disperser. The wavelength coverage ranged from 6200\AA\ to 9100\AA\ with $R = 1900$. The spectra were processed using \iraf. The white dwarf GJ 318 was used for flux calibration in May 2011, and the DA white dwarf LTT3218 was used in November 2011. The final combined spectrum is shown in Figure \ref{fig:optical_spec}. 

\subsection{X-Ray Observations}
\w2332\ was serendipitously observed twice with the ACIS-I camera in programs to map the galaxy cluster SPT-CL J2331$-$5051. The total exposure time was 35.0~ks, split between a guaranteed time observation of 29.0~ks on UT 2009 August 12 (ObsID~9333; PI Garmire) and a guest observation of 6.0~ks on UT 2009 August 30 (ObsID~11738; PI Mohr).  The X-ray data for the cluster were reported in \cite{2011ApJ...738...48A}.  The data were analyzed following standard procedures using the \Chandra\ Interactive Analysis of Observations (CIAO; V4.2) software.  We identified ``good time intervals'' for the observations, yielding a total effective exposure time of 30.0~ks for \w2332.

\section{Results and Analyses}\label{sec:results_and_analyses}

\subsection{Radio Continuum Morphology}\label{sec:radio_morph}

The ATCA maps at 1.5 GHz (20 cm), 5.6 GHz (5.4 cm), and 9.0 GHz (3.3 cm) are shown in panels (a)-(c) of Figure \ref{fig:radio_map}. These maps clearly show a mixed morphology between FR-I and FR-II structures that comprise the \w2332\ system. In the lowest resolution map, at 1.5 GHz (3\arcsec $\times$ 4\farcs8), three distinct peaks are observed, including the core and two unevenly bright spots at both ends of the center. The central peak is offset by 3\farcs0 from the location of the \wise-detected host galaxy (referred to as the core), which indicates significant unresolved 20-cm emission from the extended structure north of the core. At shorter wavelengths, the jet is well resolved and exhibits a complex morphology, described next.

\begin{figure}
\figurenum{3}
\epsscale{0.95}
\begin{center}
\plotone{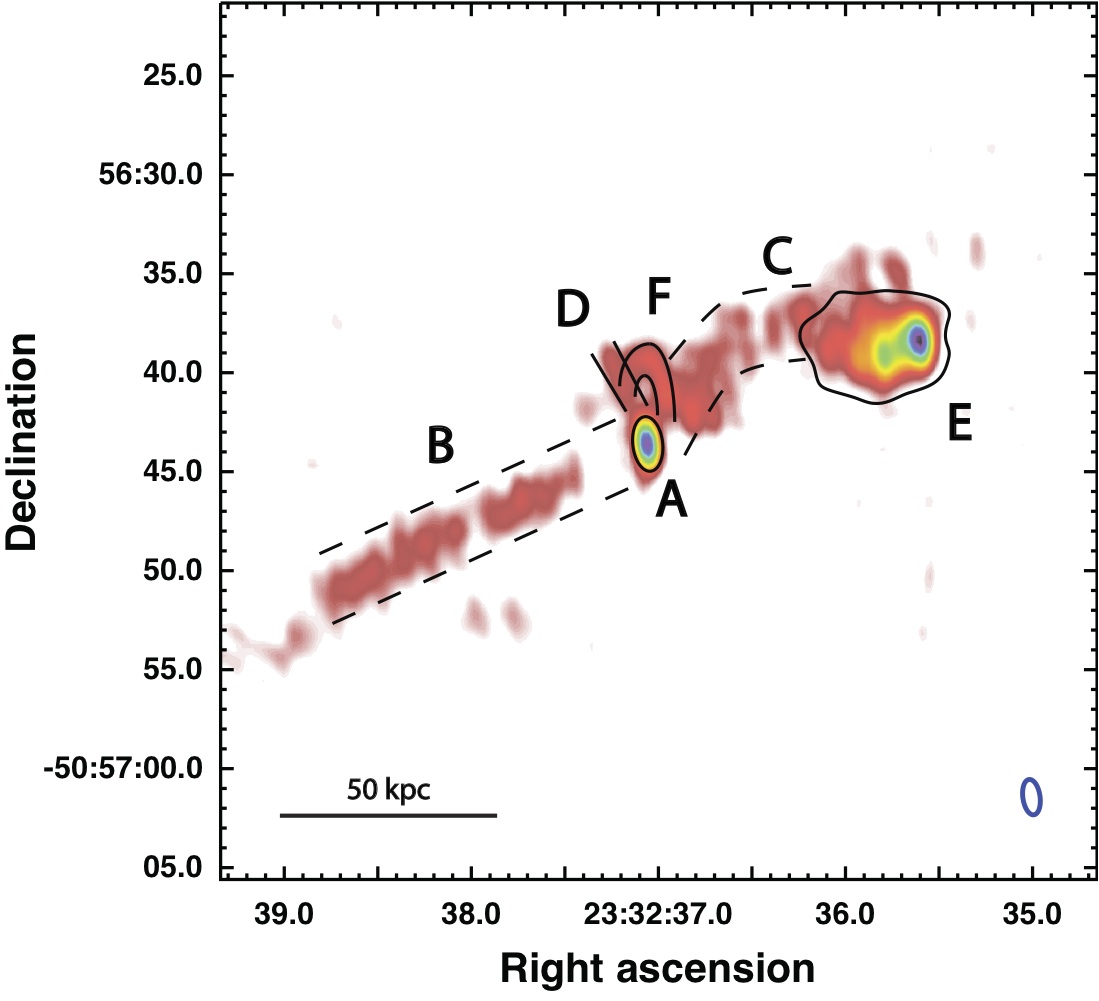}
\end{center}
\caption{9.0 GHz continuum map with the key radio features illustrated: (A) core, (B) primary south-east (SE) jet, (C) primary counter north-west (NW) jet, (D) linear structure, (E) hot spot, and (F) arc-like structure. 
}\label{fig:composite}
\end{figure}

In the 9.0 GHz map (1\farcs80 $\times$ 0\farcs87, PA = 8\degr), we identify six distinguishable components, marked as A--F in Figure \ref{fig:composite}:  A) a core (coincident with the host galaxy), a pair of ``primary'' jets, including B) a south-east (SE) jet with PA of 120\degr\ and C) the counter distorted north-west (NW) jet, D) a linear structure at a nearly orthogonal PA of 20\degr, E) a luminous hot spot, and F) an arc close to the core. The fluxes of these radio structures are listed in Table \ref{table:w2332_radio}.

\begin{deluxetable*}{lccccclll}  
\tabletypesize{\scriptsize}
\tablewidth{0in}
\tablecaption{Radio Continuum Flux Densities of Resolved Radio Components\label{table:w2332_radio}}
\tablehead{
\multicolumn{1}{c}{Component} &
\multicolumn{1}{c}{ID} &
\multicolumn{1}{c}{$F_{\rm 1.4GHz}$} &
\multicolumn{1}{c}{$F_{\rm 5.6GHz}$} &
\multicolumn{1}{c}{$F_{\rm 9.0GHz}$} &
\multicolumn{1}{c}{$\alpha^{5.6}_{9.0}$} 
\\
\colhead{} &
\colhead{} &
\colhead{(mJy)} &
\colhead{(mJy)} &
\colhead{(mJy)} &
\colhead{}
}
\startdata
Core				&	(A)	&	14$\pm$1	&	  5.0$\pm$0.1	&	  4.6$\pm$0.1					&	-0.18$\pm$0.07\\
Primary Jet (SE)	&	(B)	&	34$\pm$1	&	11.6$\pm$0.2	&	  9.5$\pm$0.3					&	-0.41$\pm$0.08\\
Primary Jet (NW)	&	(C)	&	57$\pm$1	&	27.8$\pm$0.2	&	18.1$\pm$0.3					&	-0.90$\pm$0.03\\
Linear Structure	&	(D)	&	---			&	  1.5$\pm$0.1	&	  1.1$\pm$0.1					&	-0.70$\pm$0.15\\
Hot Spot			&	(E)	&	---			&	21.2$\pm$0.1	&	14.0$\pm$0.2					&	-0.87$\pm$0.03\\
Arc-like Structure	&	(F)	&	---			&	  ---  				&	  1.83$\pm$0.08\tablenotemark{a}		&	---
\enddata
\tablecomments{Component IDs (second column) are marked in Figure \ref{fig:composite}. $F_{\rm 5.6GHz}$ and $F_{\rm 9.0GHz}$ are measured in the beam-matched maps (2\farcs9 $\times$ 1\farcs6, PA = 15\degr). The $\alpha$, as defined in $F_{\nu} \propto \nu^{\alpha}$, is the spectral index between 9.0 GHz and 5.6 GHz.}
\tablenotetext{a}{Measured in map of beam = 1\farcs80 $\times$ 0\farcs87, PA = 8\degr.}
\end{deluxetable*}  

At the position of the mid-IR and optical peak, a radio core is apparent. The core has a relatively flat spectrum: a spectral index of $\alpha = -0.18\pm0.07$ ($F_{\nu} \propto \nu^{\alpha}$) from 5.6 GHz to 9.0 GHz in the maps of matched beam of 2\farcs9 $\times$ 1\farcs6 at PA = 15\degr. The deconvolved size of the core is $<0\farcs3$ at 9.0 GHz, based on single Gaussian fitting of the uniformly weighted map.

Originating from the core, the primary jet extends about 14\arcsec\ to the north-west, punctuated by a typical FR-II galaxy Doppler-boosted hot spot that is much brighter than the core. In contrast to the linear jet (``B'') to the south-east, the north-west jet shows an asymmetric zig-zag morphology.  The deviation is 10\degr\ from the axis of the south-east jet, while the deviation of 30\degr\ is shown on the opposite side. Moreover, there appear to be distinct differences across the radio bands, notably closer to the core, better revealed in Figure \ref{fig:radio_map}(f). The jet is more curved at 5.6 GHz on the south-east side, and relatively straighter on the north-west, spatially offset from the 9.0 GHz emission. The nature of this offset is under investigation with new, deeper radio observations that we are currently pursuing.

The faint linear structure ``D'', pointing to the center of the core, extends 5\arcsec--7\arcsec\ from the central core, is visible at 5.6 GHz, and notably at 9.0 GHz (appearing as `green' in Figure \ref{fig:optical_image}(f)), extending over $>$ 2 beams. The emission at 1.5 GHz is also extended in the same direction indicated by the elongation of the core in the low resolution 1.5 GHz map. The co-alignment is shown in both 9.0 GHz and 5.6 GHz of the radio core and the resolved host in optical imaging (white/magenta in the figure). This structure has signal-to-noise $>$ 30. It is at a PA of 20\degr, which is distinct from the PA of the elongated ATCA beams at 5.6 GHz and 9.0 GHz (13\degr\ and 8\degr, respectively). Moreover, it does not match the uncleaned dirty beam structure of a relatively weak central core (peak-to-peak ratio between core and component ``D'' is $\sim$ 8). Structure ``F'' has an unusual shape resembling a curved, arc-like structure, possibly the result of a 3-dimension radio structure seen in projection. 

The primary jets extend to 70 kpc on both sides at a PA of 120\degr. To our detection limit, there is no clear jet-like structure that connects directly to the core on the axis of the primary jet in the central 12 kpc (2\farcs5). The western jet exhibits complex behavior which may indicate interactions with the interstellar medium (ISM) and intergalactic medium (IGM). The structure of the eastern counter jet is more linear, with a maximum PA change of $\la$ 10\degr. The pitch angle along the jet reaches 20\degr\ in the 5.6 GHz map, whereas it is within 5\degr\ at 9.0 GHz. If the driving mechanism that produces this ``zigzagging'' morphology in the jets is periodic, the estimated period is on the order of 0.25 Myr based on the length between off-axis peaks of the serpentine primary jet.

The asymmetric brightness of the jets suggests that the western side of the primary jet is on the side near to us. This is because jets can be brighter (fainter) by a factor of $D^{3}$ if they are moving toward (away from) the observer \citep[discussed in][]{2001Sci...291...84M}, where $D \equiv [\Gamma(1 + \frac{v}{c} \cos \theta)]^{-1}$ is the Doppler factor for a relativistic jet at Lorentz factor $\Gamma$ for given velocity $v$, and the angle between the axis of the jet and the observer's line of sight $\theta$. However, $\theta$ for the primary jets is likely to be $>$ 80\degr\ (e.g. the maximum for jets in the face-on plane) based on  the relatively small brightness difference between the primary jets on each side. 

\subsection{Spectroscopy:  Broad \Ha\ and \Hb\ Emission Lines}\label{sec:discussion_broad_line}

The integrated fluxes of identified lines are listed in the Table \ref{table:w2332_line}. The presence of the high ionization line [NeV]3426\AA, as well as the ratio of [\OIII]5007\AA\ to narrow line \Hb\ indicate the AGN nature of \w2332. Additional evidence for the AGN is provided by broad \Ha\ and \Hb\ line widths, as shown in Figure \ref{fig:optical_spec}. Both have asymmetric profiles with respect to the NL. The full width at half maximum (FWHM) of \Ha\ is $\sim$11,000 \kms, and the full width at quarter maximum (FWQM) is $\sim$ 16,000 \kms. For \Hb, the peak of the BL component is centered at $\sim$ 6453\AA, or 3800 \kms\ blueshifted with respect to the systematic host galaxy redshift of 0.3447, consistently measured by both the narrow emission lines as well as the CaHK absorption. Similarly, the peak of the BL \Ha\ is at 8730\AA, or 3200 \kms\ blueshifted with respect to the host galaxy systematic redshift. Objects with asymmetric, very broad Balmer lines whose peaks are shifted from the corresponding narrow lines are usually referred to as double-peaked emitters (DPEs). We note a few narrow emission features (marked as ``?'' in the inset of Figure \ref{fig:optical_spec}) straddling \Ha, unlikely to be [\ion{N}{2}] or [\ion{S}{2}], which could potentially be associated with \Ha\ components shifted by $-1700$ and $+4000$ \kms\ with respect to the galaxy systematic velocity. The different blueshifts for the broad \Ha\ and \Hb\ are the result of contamination from these un-identified components. From the difference between the scaled \Ha\ and \Hb\ profiles, we detect no significant broad \FeII\ emission near \Hb\ within $\pm$ 10,000 \kms. We do not see clear evidence for very broad line regions in other spectral features. 

\begin{deluxetable}{lccc}
\tabletypesize{\scriptsize}
\tablewidth{0in}
\tablecaption{Optical Line Fluxes of \w2332\label{table:w2332_line}}
\tablehead{
\multicolumn{1}{c}{Line} &
\multicolumn{1}{c}{Redshift} &
\multicolumn{1}{c}{Flux}
\\
\colhead{} &
\colhead{} &
\colhead{($10^{-16}$ erg cm$^2$ s$^{-1}$)}
}
\startdata
{[Ne\,{\sc v}]\,$\lambda 3426$}	&	0.3439		&	  7.0$\pm$0.6	\\
{[O\,{\sc ii}]\,$\lambda 3727$}	&	0.3443		&	13.4$\pm$0.4	\\
{[Ne\,{\sc iii}]\,$\lambda 3869$}	&	0.3440		&	  6.8$\pm$0.4	\\
{[O\,{\sc iii}]\,$\lambda 4363$}	&	---			&	  1.2$\pm$0.3\tablenotemark{a} 			\\
He\,{\sc ii}\,$\lambda 4686$	&	---			&	  0.8$\pm$0.3\tablenotemark{a}	\\
H$\beta$	\\
\, \,  NL					&	0.3447		&	  2.12$\pm$0.2	\\
\, \,  BL-blue				&	---			&	64.1$\pm$0.8\tablenotemark{b}	\\
\, \,  BL-red				&	---			&	71.8$\pm$1.0\tablenotemark{c}	\\
{[O\,{\sc iii}]\,$\lambda 4959$}	&	0.3449		&	11.2$\pm$0.2	\\
{[O\,{\sc iii}]\,$\lambda 5007$}	&	0.3449		&	42.2$\pm$0.2	\\
{[O\,{\sc i}]\,$\lambda 6300$} 	&	0.3458		&	  4.4$\pm$0.7	\\
H$\alpha$	 \\
\, \,  NL					&	0.3447		&	12.8$\pm$0.6	\\
\, \,  BL-blue				&	---			&    436$\pm$3\tablenotemark{b}	\\
\, \,  BL-red				&	---			&    265$\pm$2\tablenotemark{c}	\\
{[N\,{\sc ii}]\,$\lambda 6583$}	&	0.3448		&	10.5$\pm$0.6	
\enddata
\tablenotetext{a}{Marginal detection.}
\tablenotetext{b}{Integrated flux from the zero velocity to the blue end of the BL component.}
\tablenotetext{c}{Integrated flux from the zero velocity to the red end of the BL component.}
\end{deluxetable}

\subsubsection{Parameterization with Gaussian Profiles}\label{sec:spec_modeling_gaussian}
\begin{figure*}
\figurenum{4}
\begin{center}
\plotone{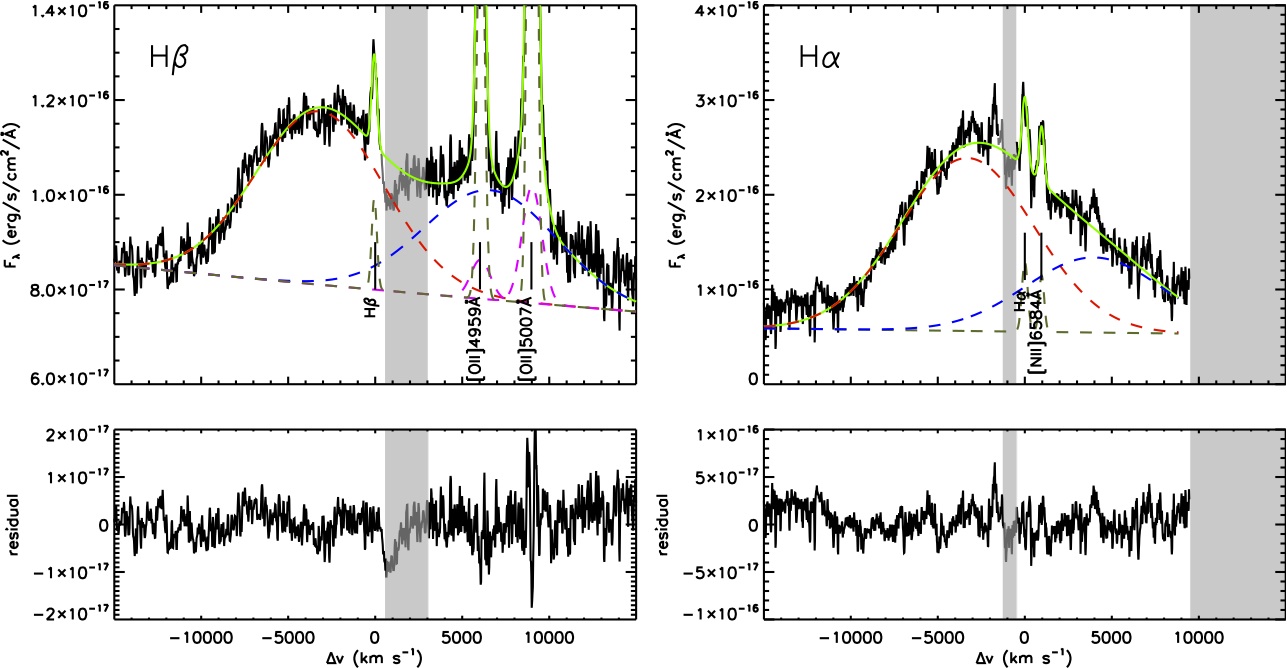}
\end{center}
\caption{Observed \Hb\ and \Ha\ broad lines overplotted with the best-fit Gaussian models of BLs plus NLs. The two Gaussian BL profiles in the model are fit independently to both of \Hb\ and \Ha. The redshift and linewidth of [\OIII]4959\AA\ and [\OIII]5007\AA\ are assumed to be equal, with a fixed flux ratio $F_{\rm [OIII]5007}/F_{\rm [OIII]4959} = 3$. The shaded regions which cover the narrow absorption features and wavelengths outside the spectral window are excluded in the model fitting. The blue and red dashed lines represent the blueshifted and redshifted BL components of \Ha\ and \Hb, respectively. The magenta lines indicate the BL components of  [\OIII]4959\AA\ and [\OIII]5007\AA. The dark green dotted lines show NL components of \Hb, [\OIII]4959\AA, [\OIII]5007\AA, \Ha, and [\NII]6584\AA\ lines. The green solid line is the sum of all components. The lower two panels show the residuals from the fitting.}\label{fig:spec_gaussian}
\end{figure*}

We first attempted to parameterize the BL components with two Gaussian profiles without excluding narrow emission lines in order to measure the FHWM and fluxes of all lines. The two BL Gaussian profiles and one NL component are used for both \Ha\ and \Hb\ in the model, allowing to fit independently on \Hb\ and \Ha. The narrow absorption features (between rest-frame 4866.4--4908.2\AA\ and 6529.3--6552.4\AA) are excluded from the fitting process. For the [\OIII]4959\AA\ and 5007\AA\ lines, one BL Gaussian component and one NL component are assumed, while the [\NII]6583\AA is fit to a single NL Gaussian component. The redshift and linewidth of the [\OIII]4959\AA\ and [\OIII]5007\AA\ lines are assumed to be equal, with a fixed flux ratio $F_{\rm [OIII]5007}/F_{\rm [OIII]4959} = 3$. The results ($\chi^{2}_{\rm{red}}=1.85$, dof $=3607$) are shown in Figure \ref{fig:spec_gaussian} and Table \ref{table:lines_gaussian}. 

The blue BL components are centered at $\lesssim -4300$ \kms\ with respect to the narrow line, while the red component peaks at $> +5300$ \kms. Both blue and red BL components have FWHM of $\gtrsim$ 8800 \kms. The flux of the blue components is $\sim$ 1.5--2.3 times that of the red components. The fitting residuals shown in the lower panels of Figure \ref{fig:spec_gaussian} indicate that the model represents the data near both \Ha\ and \Hb\ fairly well, except the blocked region and the very red wing of the \Hb\ ($\Delta V > 11,000$ \kms).

\begin{deluxetable}{lrrcc}
\tabletypesize{\scriptsize}
\tablewidth{0in}
\tablecaption{Emission Line Parameters\label{table:lines_gaussian}}
\tablehead{
\multicolumn{1}{c}{Line} &
\multicolumn{1}{c}{$\Delta V$} &
\multicolumn{1}{c}{FWHM} &
\multicolumn{1}{c}{Flux}
\\
\colhead{} &
\colhead{(\kms)} &
\colhead{(\kms)} &
\colhead{($10^{-16}$ erg cm$^2$ s$^{-1}$)}
}
\startdata
H$\alpha$	 \\
\, \,  NL					&	$0\pm$40			&	390$\pm$50		&	9$\pm$1	\\
\, \,  BL-blue				&	$-4410\pm$50		&    	9180$\pm$30		&	519$\pm$2\\
\, \,  BL-red				&	$+5330\pm$130	&    	9110$\pm$60		&	229$\pm$3\\
H$\beta$ \\ 
\, \,  NL					&	$-70\pm$30		&	360$\pm$60		& 	1.6$\pm$0.2	\\
\, \,  BL-blue				&	$-4270\pm$90		&	8790$\pm$10		&	74$\pm$1	\\
\, \,  BL-red				&	$+8860\pm$130	&	8800$\pm$20		&	48$\pm$1	\\
{[O\,{\sc iii}]\,$\lambda 5007$}\\
\, \,  NL					&	$+70\pm$10		&	480$\pm$10		&	33$\pm$1	\\
\, \,  BL					&	$+50\pm$50		&   	1300$\pm$120		&	8$\pm$1	\\
{[N\,{\sc ii}]\,$\lambda 6583$}\\
\, \,  NL					&	$-10\pm$10		&	340$\pm$60		&	6$\pm$1
\enddata
\tablecomments{The results of Gaussian fitting of the emission lines are shown here. The second column, $\Delta V$, represent the velocity offset with respect to the rest frame wavelength of the line at redshift of 0.3447. For each emission line, one BL component and one NL component is included, except \Ha\ and \Hb\ for which two BL components are employed. The [O\,{\sc iii}]4959\AA\ (not shown in the table) and [O\,{\sc iii}]5007\AA\ lines are forced to have the same $\Delta V$ and FWHM, with a 1:3 line flux ratio.}
\end{deluxetable}

\subsubsection{Spectroscopic Modeling -- the Case of a Thin, Elliptical Accretion Disk}\label{sec:spec_modeling}

In the literature, the DPE feature, such as we see in \w2332, has often been interpreted with a thin accretion disk surrounding a single AGN \citep{2007ApJ...666L..13B,2009ApJ...696.1367S,2009NewAR..53..133E}. We test the disk-emitter scenario by fitting the \Ha\ and \Hb\ profiles with models based on line emission from an elliptical accretion disk \citep{1995ApJ...438..610E}. The asymmetric blueshifted broad line features in both \Ha\ and \Hb\ in this model are due to Doppler boosted line emission from the disk. All NL features (\Ha, \Hb, [\OIII]4959\AA, [\OIII]5007\AA, [\NII]6584\AA) and an unexpected `absorption-like' region redward of NL \Hb\ (6535 -- 6600\AA) are excluded from the model fitting. The eccentricity, $e$, angle of periastron, $\phi$, inclination angle, $i$, and inner/outer radii, $r_{\rm in}$/$r_{\rm out}$, are free parameters for fitting. Each model fit \Ha\ and \Hb\ with a radius-dependent power-law emissivity $\epsilon(r) = \epsilon_0 \times r^{-q}$ for the disk, where $r$ is the radius of the emission region from the center, and $q$ is the power law index. Two $q$ indices are used as independent parameters, $q_{\rm H\alpha}$ and $q_{\rm H\beta}$, for the emissivities of \Ha\ and \Hb, respectively. Theses emissivity assumptions for the photoionization models \citep{1989A&A...213...29C,1990A&A...229..313D,1990A&AS...83...71D} were developed by \cite{1989ApJ...344..115C}, and are extensively used by others, including \cite{2003ApJ...599..886E}.

\begin{figure*}
\figurenum{5}
\begin{center}
\plotone{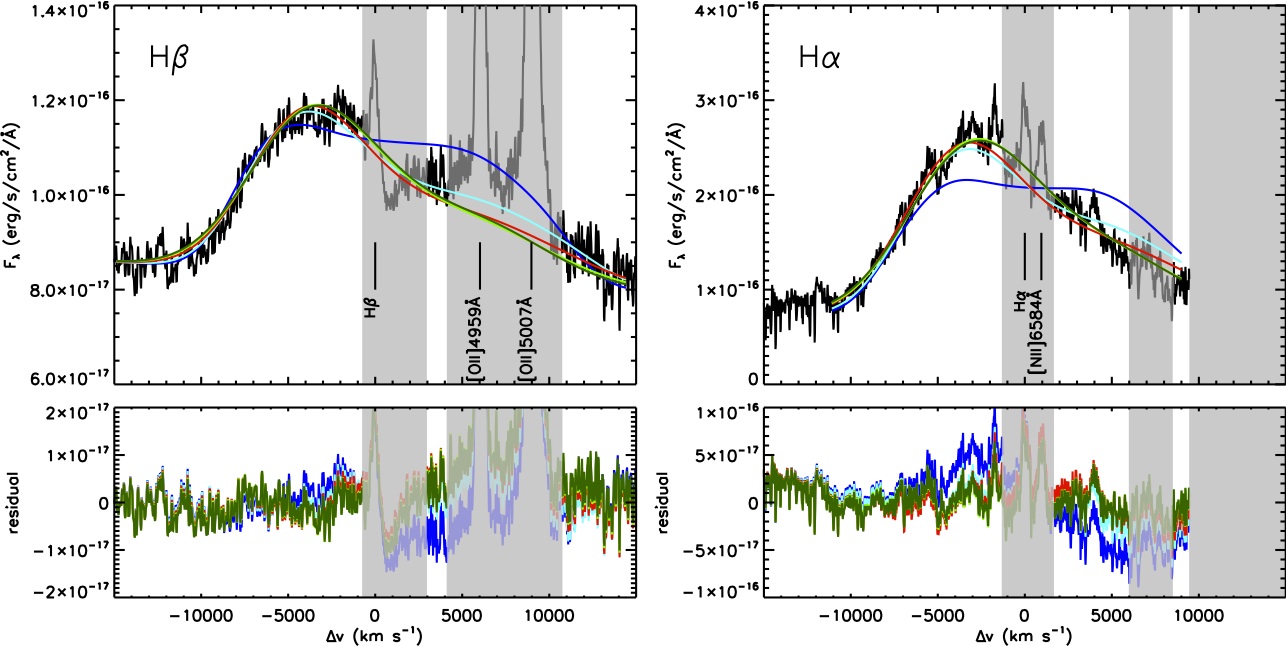}
\end{center}
\caption{Observed \Hb\ and \Ha\ broad lines overplotted with elliptical disk models for five eccentricities with all other parameters the same: $e=0.0$ (blue), $e=0.2$ (cyan), $e=0.4$ (red), $e=0.6$ (green) and $e=0.8$ (dark green). Models are fit simultaneously on both \Hb\ and \Ha\ with an assumption of a radius-dependent power-law emissivity. See text for details. The shaded regions are excluded in the model fitting. The $e=0.6$ model yields the best fit, though the broad line profiles can be fit relatively well by the $e=0.2$, 0.4, and 0.8 models.}\label{fig:spec_elliptic_ind_disk}
\end{figure*}

The modeling was performed by fitting a grid of free parameters and adopting the set of parameters that yields the lowest reduced chi-squared value. We tested elliptical disk models for eccentricities of $e=$ 0.0 -- 0.8 (shown in Figure \ref{fig:spec_elliptic_ind_disk} along with the best fit circular model). The high eccentricity $e=0.6$ model provides a statistically significant improvement to explain the BL profiles ($\chi^{2}_{\rm{red}}=0.88$, dof $=1887$), although models with $e >$ 0.2 generally provide good fit ($\chi^{2}_{\rm{red}}\leq1.02$). The best-fit parameters of this model are as follows: $e=0.6$, $q_{\rm H\alpha} = 2.4$, $q_{\rm H\beta} = 3.2$, $i=42^{\circ}$, FWHM local broadening $\sigma=3300$ \kms, $r_{\rm disk}^{\rm inner}=240~r_{\rm G}$, and $r_{\rm disk}^{\rm outer}=4500~r_{\rm G}$, where $r_{\rm G}$ is the Schwarzschild radius of the central black hole. 

This model with a thin disk and offset BL regions can reasonably explain the \Ha\ and \Hb\ line profiles in \w2332, other than an apparent ``absorption'' feature at $\Delta V \sim$ 1000 \kms\ redshifted with respect to \Hb. This absorption-like profile is not observed for \Ha, likely because it overlaps with [N{\sc ii}]6583\AA\ emission at the systematic redshift of the galaxy. The nature of this feature is yet unknown. 

\subsection{X-ray Emission}\label{sec:xray}

The AGN hosted by \w2332\ is strongly detected in {\it Chandra} ACIS-I data, with a total of 1790 source counts in the energy range $0.3 -7$\,keV. It is the brightest X-ray source in the field. In the deeper observation (ObsID 9333), our target is 9\farcm3 off-axis, where the \Chandra/ACIS 50\% encircled energy radius is 4\arcsec\ at 1.4\,keV. We extracted the source distorted by the off-axis PSF using a relatively large 12\arcsec\ radius circle. Background was from a 30\arcsec\ circular source free region. Response matrices and effective areas were then determined for the target position, and the spectrum was binned to have at least 20 counts per spectral bin. The resulting spectrum is analyzed using XSPEC (V12.6.0).   

\w2332\ appears spatially extended in the \Chandra\ data, but the distorted off-axis PSF complicates the interpretation of this spatial structure. We used the CIAO task \textsc{arestore} to perform PSF deconvolution on the deeper image (ObsID 9333) with a PSF simulated by \textsc{CHart} in the energy range of 0.3--7 KeV. Two weak, extended X-ray regions $\sim$ 4\farcs5 from the central X-ray peak at PA of $\sim$ 30\degr\ and $\sim$ 200\degr\ are marginally detected; together they contains $\sim 0.6\%$ of the X-ray photon counts. 

Using the standard definition of hardness ratio, $HR \equiv (H-S)/(H+S)$, where $H$ is the hard counts between $2-10$\,keV and $S$ is the soft counts between $0.5-2$\,keV, the derived $HR = -0.05$ is typical of extragalactic X-ray sources \citep[e.g.,][]{2002AJ....123.2223S}.  Our best absorbed power-law fit ($\chi^2_{red}=1.09/70$dof) to the spectrum has an observed photon index $\Gamma = 1.51 \pm 0.08$, a total absorbing column of $N_{\rm H} = 5.31^{+0.92}_{-0.86}\, \times 10^{21}\, {\rm cm}^{-2}$, an observed soft X-ray flux of $f(0.5-2\, {\rm keV}) =2.32^{+0.34}_{-0.32}\, \times 10^{-13}\, {\rm erg}\, {\rm cm}^{-2}\,{\rm s}^{-1}$, and an observed hard X-ray flux of $f(2-10\, {\rm keV})= 1.01^{+0.20}_{-0.17}\, \times 10^{-12}\, {\rm erg}\, {\rm cm}^{-2}\,{\rm s}^{-1}$. The corresponding intrinsic, or unobscured X-ray luminosities are $L(0.5-2\, {\rm keV}) = 1.72^{+0.06}_{-0.06}\, \times 10^{44}\,{\rm erg}\, {\rm s}^{-1}$ and $L(2-10\, {\rm keV}) = 4.14^{+0.32}_{-0.28}\,\times 10^{44}\, {\rm erg}\, {\rm s}^{-1}$.

\subsection{Variability}\label{sec:variability}

\w2332\ was observed in {\it Spitzer} [3.6] in 2009 and in WISE [3.4] between 2010--2011. The median of the [3.4] single-frame fluxes is 1.30$\pm$0.05 mJy while the [3.6] flux from a single measurement is 1.311$\pm$0.002 mJy; the anticipated color difference between IRAC [3.6] - WISE [3.4] is less than 0.1 magnitude\footnote{based on Figure 4 of \url{http://wise2.ipac.caltech.edu/docs/release/allsky/expsup/sec6_3a.html}}. No significant mid-IR variation is seen. 

\begin{figure*}
\figurenum{6}
\begin{center}
\plotone{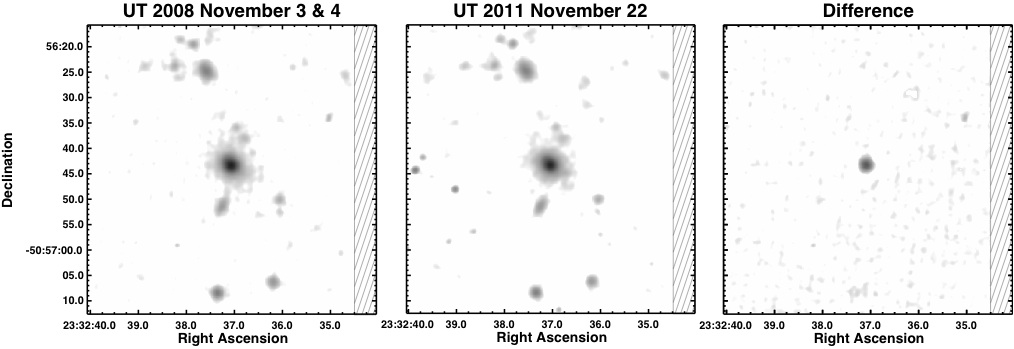}
\end{center}
\caption{
Optical $r^\prime$ images of \w2332\ from two different epochs (left two panels) after convolving to a matched $1\farcs2$ FWHM PSF, and the residual difference image (right panel). The shaded region illustrates the region in the CCD gap in the 2011 observation. This region is excluded in the analysis. All 5 $\sigma$ sources, including the extended structure of \w2332, are successfully subtracted down to the $< 2\, \sigma$ level. The residual component of \w2332, coincident with the nucleus, has a FWHM of $1\farcs2$, indicating that it is unresolved. The peak flux of the residual is $\sim 140\, \sigma$ of the difference image, and contains about 18\% of the flux of the peak in November 2008 which is brighter than that in 2011.
}\label{fig:optical_variation}
\end{figure*}

At optical wavelengths, the redshifted \Hb\ spectra were obtained twice in 2011, separated by 6 months. There is no obvious variation in \Hb\ between these two epochs. On longer time spans ($>$3 years) the central AGN does show significant fluctuations at optical wavelengths. The residual in the differential $r^\prime$-band imaging (Figure \ref{fig:optical_variation}) indicates a $\sim$ 20\% total flux decrease from 2008 November to 2011 November.

\subsection{Optical Imaging Morphology}\label{sec:morphology}

\begin{figure*}
\figurenum{7}
\begin{center}
\plotone{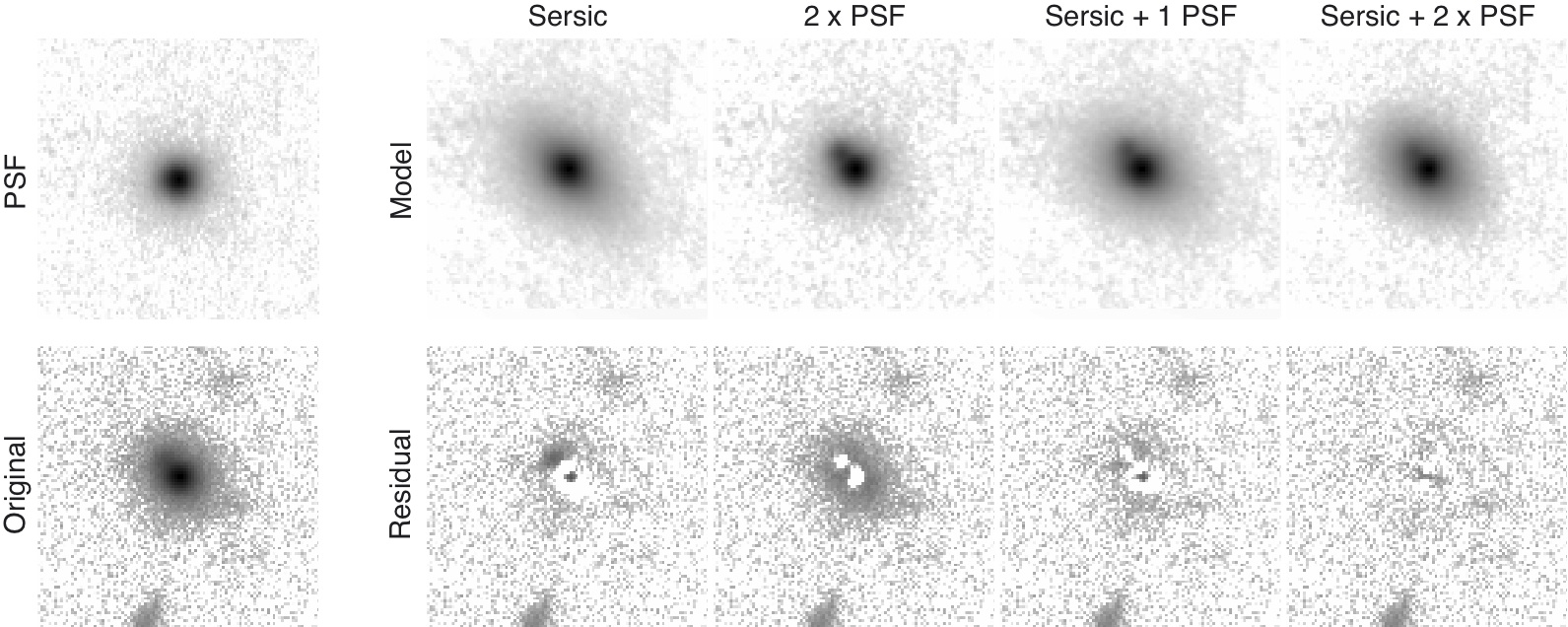}
\end{center}
\caption{GALFIT morphological models and residuals images of \w2332\ in $r^\prime$ band. The PSF and the original image (taken on UT 2012 November 22 with seeing of 0\farcs65) are shown on the left. The other eight panels, in square-root scale,  show models and residual images (14\arcsec\ on a side). Note that the free single PSF component in the ``\sersic\ + 1 PSF'' model (the second column from the right) selects an off-center optical bright spot instead of the central peak.}\label{fig:galfit}
\end{figure*}

\w2332\ is clearly resolved in the SOAR $r^\prime$-band image (see Figure \ref{fig:radio_map}e and Figure \ref{fig:galfit}; 0\farcs65 seeing). We use GALFIT 3.0 \citep{2002AJ....124..266P,2010AJ....139.2097P} to examine its morphology. Inspired by evidence of at least one BL AGN in the system, we use a combination of \sersic\ profiles and PSFs to model the host galaxy and unresolved AGN components, respectively. The results are shown in the right panels of Figure \ref{fig:galfit}. The single \sersic\ model, single PSF model, and dual PSF models can not explain the observed morphology well; all of them leave significant residual flux ($>$ 3\%) within 4\farcs8 radius of the core. Modeling the image with a \sersic\ profile plus a single free PSF, the PSF preferentially selects an off-center optical bright spot, but leaves significant central residual flux ($\sim$ 7\%). The best fit model ($\chi_{\rm red}^{2} \sim 1.4$, dof = 40388) requires a single \sersic\ profile accompanied with two PSF profiles which match the observed brightness distribution to $<$ 0.5\%. In this model, the \sersic\ component, primary PSF component, and secondary PSF components contribute 33.7\%, 61.8\%, and 4.5\% of the total flux, respectively. 

\begin{figure*}
\figurenum{8}
\epsscale{0.75}
\begin{center}
\plotone{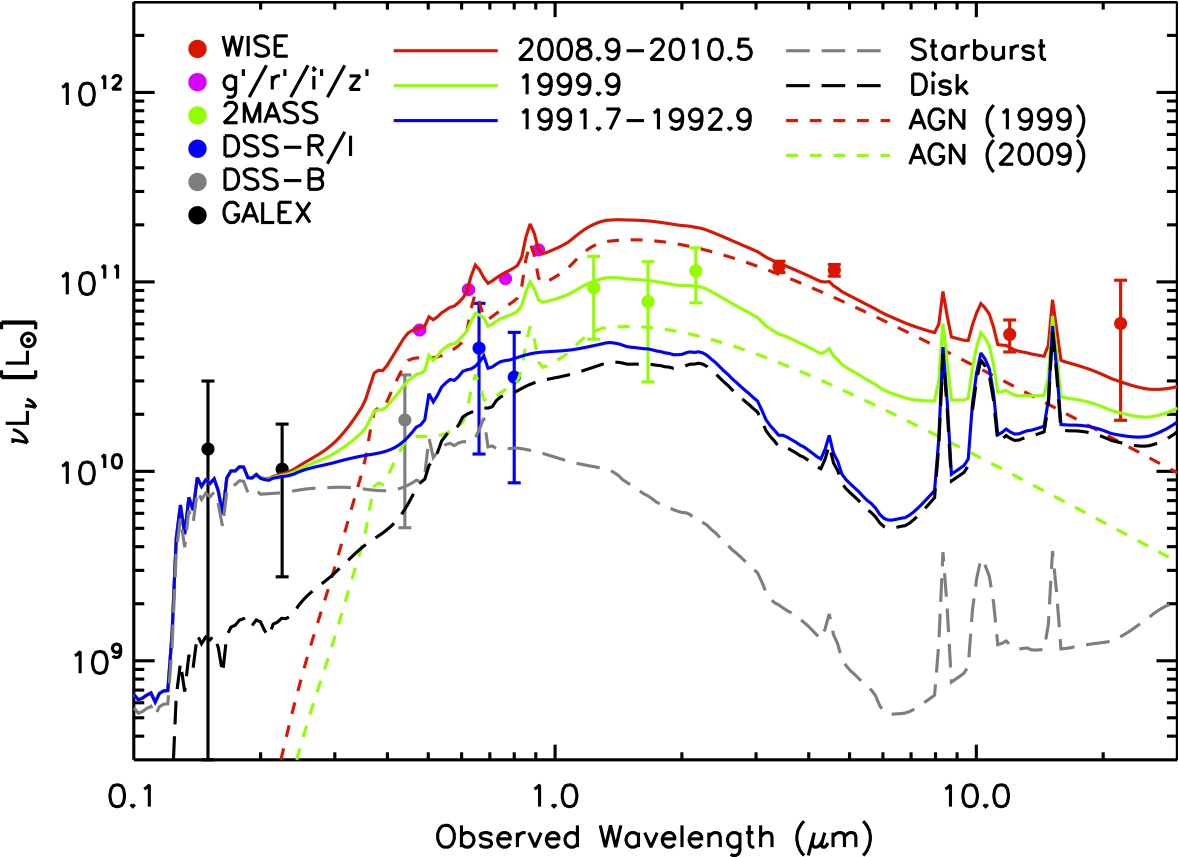}
\end{center}
\caption{UV-IR SED models of \w2332. The dashed lines represent the different components in the SED, including starburst (grey), normal disk (black), and AGN (red and green), based on the empirical SED templates from \cite{2010ApJ...713..970A}. Because of the variability between the different epochs (see \S \ref{sec:variability}), we separate the fitting into three different epochs: 1991 August - 1992 November (\GALEX\ + DSS -- blue), 1999 October (\GALEX\ + 2MASS -- green), and 2008 November -- 2010 May (\GALEX\ + \textit{g$^\prime$r$^\prime$i$^\prime$z$^\prime$} + \WISE\ -- red). The DSS-$B$ data (grey dot) taken in 1975 is not included in the fitting. The UV data from \GALEX\ is fixed for fitting all epochs (see text for details). The earliest epoch data (blue solid line) is well fit by a galaxy without any AGN component. The later two epochs (green and red solid lines) are fit by adding an AGN component which brightens over time.}\label{fig:sed}
\end{figure*}

The \sersic\ profile (of the host galaxy) in the best-fitted model has an axis ratio of 0.64 at a PA of 48\degr. It contributes $\sim$ 50 $\mu$Jy (or AB = 19.64) with an effective radius of 1\farcs36 (6.7 kpc), and has a \sersic\ index of 0.92, close to an exponential disk index $n=1$. This component has $M_{\rm r^{\prime}} = -21.6$ (AB mag) before $k$-correction, slightly brighter than the Milky Way. The luminosity is $2.5 \times 10^{10}~ L_{\sun}$, similar to our SED model of the host galaxy (blue line in Figure \ref{fig:sed}). The primary PSF component is centered 0\farcs16 offset ($\sim$ 800 pc in projection) from the \sersic\ center. This displacement, 25\% of the PSF FWHM, appears significant, but could be the result of an imperfect PSF generated from unresolved sources in the field.

The secondary unresolved component is 1\farcs0 (4.9 kpc in projection) offset from the \sersic\ center. This component does not show obvious variability. It is too bright ($M_{\rm r^{\prime}} = -19.4$) to be a star cluster inside the host galaxy or a satellite dwarf elliptical galaxy, though it could be a second merging, compact galaxy. It is not likely to be stellar object unless it is a foreground star along the line of sight. The position angle to the primary compact object is 49\degr, or $\sim$ 30\degr\ with respect to the secondary radio jet, making it unlikely to be a bright spot associated with an optical jet of the central AGN. Further understanding of this component will require sensitive, higher resolution imaging.

The brightest visual companion at slightly greater disance ($r >$ 4\arcsec, $\sim$ 20 kpc) has $m_{\rm r^{\prime}}$ is $21.6$ (AB mag), which corresponds to $\sim 6\times 10^9 L_{\sun}$ in $r^{\prime}$ band after $k$-correction assuming the redshift of \w2332. There are no obvious disturbed morphological features brighter than 1/5 of the luminosity of \w2332\ to suggest that the host galaxy is an early/intermediate-stage major merger system.

\subsection{SED Modeling}
\label{sec:sed}

We model the broad-band SED of \w2332 from the UV to the mid-IR with a combination of host galaxy and AGN components using the empirical SED templates of \cite{2010ApJ...713..970A}.  One challenge is that the source is clearly variable at optical wavelengths (see \S \ref{sec:variability} and Figure \ref{fig:optical_variation}). Indeed, the red optical SED observed by Magellan in 2008 is difficult to reconcile with the very blue $z^{\prime} - J$ color implied by the 2MASS $J$-band photometry obtained in 1999 unless we invoke variability.  We therefore divide the photometric data into three epochs (1991.7--1992.9, 1999.9, and 2008.9--2010.5, in roughly 3-year-long time bins) for SED fitting purposes (Figure \ref{fig:sed}).

The multi-epoch SED fitting was done requiring a common, non-variable host galaxy using a combination of empirical starburst, disk galaxy, and elliptical galaxy templates from \cite{2010ApJ...713..970A} to represent the different age stellar populations in \w2332.  We required these SED galaxy components to match the morphological modeling discussed in \S \ref{sec:morphology}; namely, we required that the $r^{\prime}$ fluxes from the three galaxy components sum to one-third of the total $r^{\prime}$ flux, or $\sim 60 \mu$Jy.  For the AGN component, we adopted a modified AGN SED template in which the warm dust emission at $\lambda > 1\micron$ is replaced by a power-law emission to continue the accretion disk emission to longer wavelengths. This modification was made in order to satisfy a reasonable fit between $r^{\prime}$ band and W2. We also apply foreground dust extinction toward the AGN to accommodate the red optical color. This modified obscured AGN component is reminiscent of the synchrotron emission that characterizes blazars at these wavelengths \citep{2010ApJ...716...30A}; the WISE colors of \w2332\ put it close to the blazar strip of \cite{2012ApJ...750..138M}. Blazars, conventionally considered to be the radio galaxies with relativistic jets close to the line to sight, exhibit large amplitude luminosity variations over hours to days \citep[see review by][]{1997ARA&A..35..445U}.  However, the interpretation of \w2332\ as a blazar is inconsistent with the radio morphology, nor do we see strong evidence for short-time scale variability at optical and mid-IR wavelengths. We do not address this blazar interpretation further here.

The results of the model fitting are presented in Figure \ref{fig:sed}. The \GALEX\ observations, obtained between 2003 and 2005, would appear to present a challenge. On the one hand, they could be associated with a variable AGN component observed during an epoch when we do not have other photometry available to use in the SED modeling. On the other hand, they could be associated with the non-variable host galaxy, or a combination of the host and variable AGN. In practice, our results do not depend strongly on the \GALEX\ data.  If we completely ignore those UV data, our best fit model implies a heavily obscured AGN during the latest epoch with $E(B-V) = 0.81 \pm 0.05$.  For this model, the host galaxy, dominated by the starburst (e.g., irregular galaxy) component, goes through the \GALEX\ photometric points with the best fit luminosity to the irregular component being $L_{\rm im} = (2.6 \pm 1.5)\, \times\, 10^{10}\, L_\odot$. Additionally, the $\chi_{red}^{2}$ increases from 1.65 (dof = 6) to 3.26 (dof = 7) if we exclude the irregular component in the SED fitting. These suggest that the \GALEX\ photometry corresponds to the host galaxy with little AGN contribution. If we redo the fitting under that assumption, the resulting AGN obscuration basically remains the same, $E(B-V) = 0.82 \pm 0.04$, and the starburst component luminosity becomes slightly more robust, but is essentially unchanged, $L_{\rm im} = (3.2 \pm 1.0)\, \times\, 10^{10}\, L_\odot$. It is this latter case where we assume the decade-old \GALEX\ photometry is due to the host galaxy which we show in Figure \ref{fig:sed}. 

An obscured AGN dominates the optical and WISE photometry for the most recent epoch, while ten years ago, when the 2MASS photometry was obtained, our modeling is based on fewer bands but suggests a slightly less luminous AGN with a similar level of obscuration, $E(B-V) = 0.8$.  The AGN reddening in these epochs is slightly higher than the $E(B-V)=0.4\pm0.1$ implied from the X-ray column, $N_{\rm H} \sim 2.3\pm0.5 \times 10^{21}\, {\rm cm}^{-2}$ (\S \ref{sec:xray}), if the conventional Milky Way value of $N_{H}/E(B-V) \sim 5.9 \times10^{21}\, {\rm cm}^{-2}$ \citep{1978ApJ...224..132B,2002ApJ...577..221R} applies. Finally, the same fitting was applied to the DSS photometry for epoch 1991.7--1992.9. Remarkably, the best-fit model shows no AGN emission component is necessary, and is relatively well modeled by the host galaxy component alone. The DSS $B$-band, taken in 1975, is not included in the fitting due to its early observation epoch, but appears not to affect the fitting results if included. The best-fit AGN amplitude is nominally zero, suggesting that only $\sim$ 20 years in the past \w2332\ did not have appreciable nuclear activity. We note that, although the best fit model has relatively larger uncertainty due to the quality of the DSS photometry, the strong variation in the optical band over two decades is significant. The interpolation of our high-quality optical photometry measurements in the current epoch (2008--2010) suggests a flux of $\sim 300$\,$\mu$Jy in the DSS $I$-band measurement, $> 10 \sigma$ higher than the measurement of $80\pm20$\,$\mu$Jy in 1991--1992 epoch.

\subsection{Host Galaxy -- a PRONGS Candidate}

FR-II radio sources are almost invariably found to be associated with large elliptical galaxies \citep{1964ApJ...140...35M,1989MNRAS.238..357O}. Such hosts typically have de Vaucouleurs' $r^{1/4}$ profiles \citep{1991MNRAS.249..164O}, corresponding to a \sersic\ index of 4. However, a class of radio sources hosted in star forming galaxies has been identified by \cite{2006AJ....132.2409N}, and later named as Powerful Radio Objects Nested in Galaxies with Star formation \citep[PRONGS;][]{2010IAUS..267..119M,2012MNRAS.422.1453N}. These objects are radio-loud AGNs ($\ga 10^{24}$ W\,Hz$^{-1}$) with radio lobes $<$ 200 kpc that are hosted by spiral or starburst galaxies. \cite{2010IAUS..267..119M} discovered 50 PRONGS in a systematic search of a 7 deg$^{2}$ field in the Australia Telescope Large Area Survey \citep[ATLAS;][]{2006AJ....132.2409N,2008AJ....135.1276M}. Prior to this work, two radio-loud edge-on disk galaxies (NGC\,612 and 0313-192) were known in local Universe \citep{1978A&A....69L..21E,2001ApJ...552..120L,2006AJ....132.2233K,2008MNRAS.387..197E}. NGC\,612 has a collimated FR-I-like jet, and an FR-II-like lobe with a bright hot spot on the other end, sometimes described as a hybrid-morphology radio source \citep[HYMORS;][]{2000A&A...363..507G}, while 0313-192 has a typical FR-I morphology. There are a few other similar nearby examples where radio-loud objects are hosted in disk-like galaxies such as 3C305 \citep{1982ApJ...262..529H}, 3C293 \citep{1984ApJ...277...82V}, B2 0722+30 \citep{2009MNRAS.396.1522E}, PKS 1814-636 \citep{2011A&A...535A..97M}, and SDSS J1409-0302 \citep{2011MNRAS.417L..36H}. They could be associated with PRONGS but perhaps at a slightly different evolutionary phase. Previous work has suggested that PRONGS may represent an early evolutionary stage for radio-loud galaxies. An example is given by \cite{2012MNRAS.422.1453N}, who present a VLBI image of a highly embedded, compact core-jet AGN with 1.7 kpc-long jets in IRAS F00183-7111, an ULIRG with $L_{\rm bol} \sim 9 \times 10^{12}\,L_{\sun}$ and substantial star formation activity. 

The optical morphology (\S \ref{sec:morphology}) suggests that the host galaxy of \w2332\ has a modest disk (\sersic\ index $\sim$ 0.9), not much bigger than the Milky Way in size and luminosity. With a 5GHz-to-4400\AA-continuum flux density ratio $\gg$ 10, the \w2332\ system is clearly hosting a radio-loud AGN according to the definition of \cite{1989AJ.....98.1195K}. Although hosting a HYMORS-like source, the galaxy of the \w2332 system seems to be a normal disk galaxy, making it a candidate for the rare PRONGS classification. 

Based on its rest-frame NUV luminosity $\nu L_{\nu} \sim 1.1 \times 10^{10}\, L_{\sun}$ from the host galaxy SED, we estimate a star formation rate of 5 $M_{\sun}\,{\rm yr}^{-1}$ in \w2332, using Equation 1 of \cite{1998ARA&A..36..189K} with no significant dust attenuation assumed. The total luminosity of \w2332\ up to 22 \micron\ is $3.7 \times 10^{11}\, L_{\sun}$. The non-detection by {\it AKARI}, {\it IRAS}, and SPT limit the FIR luminosity ($L_{\rm FIR}$) to $3.5 \times 10^{12}\,L_{\sun}$ if an Arp220-like FIR SED is assumed. However, this Arp220-like SED model does not match the observed optical-to-near-IR photometry of the system. If a fraction of the star formation activity is highly obscured at optical wavelengths ($A_V > 40$) but releasing energy via dust emission, the 22\micron\ (W4) flux density limits $L_{\rm FIR}$ to $< 3.5 \times 10^{11}\,L_{\sun}$, or $< 30\,M_{\sun}\,{\rm yr}^{-1}$ \citep[e.g., 10\% that of Arp220;][]{2000ApJ...537..613A}. Thus, the host galaxy of \w2332 is not likely to be undergoing a major starburst. This star formation rate is close to a typical quiescent disk galaxy, unlike the hosts of other PRONGS. 

section{Discussion}\label{sec:discussion}

Observationally, as discussed in \S\ref{sec:results_and_analyses}, \w2332\ shows (i) unusual, complex radio morphology, including curved jets, and an emission complex within the central 25 kpc, (ii) a disk-like galaxy as the host of an FR-I/FR-II hybrid radio source, (iii) long-term (3--20 yrs) optical-near-IR nuclear variability, (iv) DPE profile in \Hb\ and \Ha. The standard interpretation of DPEs in which the BLs originate from the surface of a geometrically thin accretion disk around a single central SMBH is not ruled out by the observed optical spectrum, and do not prohibit the observed photometric variability. Our best-fit single AGN model (\S \ref{sec:spec_modeling}) entails an inclined elliptical ($e = 0.6$) thin disk at a viewing angle of $\sim$ 42\degr\ from the face-on. The lifetime of high eccentricity could be a concern, though the circularization of the disk is from inside-out; thus the outer disk can retain its ellipticity for longer periods of time \citep{1992MNRAS.255...92S}. On the other hand, a warped disk, or a single-arm spiral disk can mimic the BL profile similar to that from elliptical thin disk. These possibilities can be examined through the BL profile variability from a long term spectroscopic monitoring, and detailed line profile modeling, although there is no detectable variation in the line profile over the $\sim$ 6 month period separating the two GMOS observations.

The radio emission features, however, indicate that \w2332 has a more complicated configuration than a typical disk-emitter. First, the viewing angles of the primary jets and the accretion disk in the thin disk model are not consistent with each other. The flux ratio of the two primary radio jets along the inner 4\farcs5--9\farcs5 is $\sim 2$. This fairly small ratio suggests the primary jets are close to be in the face-on plane (the angle between the jet flow direction and the line of sight $>$ 80\degr) unless the two jets are intrinsically different. The inclination of the best-fitted elliptical disk, however, is 30\degr--50\degr\ after relaxing other fitting parameters such as eccentricity to $e = $ 0.2 -- 0.8. Secondly, the appearance of the radio emission complex (such as component ``D'' and ``F'' in Figure \ref{fig:composite}) within projected 25 kpc from the core suggests some event or continuous activities within the past 70,000 years if the complex originated from the core and propagates relativistically. Since the complex has high radio luminosity ($\sim 10^{40}\ {\rm erg}\,{\rm s}^{-1}$), and it does not have optical or mid-IR counterparts, the assumption that it originates from the AGN is viable. The causes of the radio complex could easily create imprints (such as a hot spot or a gap) in the geometric configuration of the thin disk, if the disk configuration is not part of the cause itself.

We should note that if a rotating BH is misaligned with the accretion disk, the material will be realigned due to Lense-Thirring drag \citep{1975ApJ...195L..65B}. SMBH-accretion disk realignment, which is sometime attributed to the cause of the XRGs (see \S \ref{sec:intro}), has a timescale below Myr for a SMBH of $10^{8}\,M_{\sun}$ under typical AGN accretion rates \citep{2002MNRAS.330..609D}. We crudely estimate the SMBH mass of \w2332 (with radio luminosity of $\nu L_\nu = 5.7 \times 10^{40} \,{\rm erg\,s^{-1}}$ at 5.6 GHz and a core X-ray luminosity of $L(2-10\, {\rm keV}) = 4.1 \times 10^{44}\ {\rm erg}\,{\rm s}^{-1}$) to be $\sim 4 \times 10^{8}\,M_{\sun}$ using the fundamental plane of black hole accretion \citep{2003MNRAS.345.1057M,2009ApJ...706..404G}. The uncertainty on this mass is a factor of $\sim$ 8 due to the scatter of the empirical relations between X-ray luminosity, radio luminosity, and BH mass. The large uncertainty in the SMBH mass will not affect the realignment timescale too much since it has a low dependency ($\propto M^{-1/16}$) to the BH mass. It is viable to consider that the radio complex is the result of jet reorientation due to the BH-disk realignment in the past 0.1 Myr, although the cause of the initial misalignment between SMBH spin and the accretion disk is unknown. Under this scenario, the double-peaked BL profile is produced by the warped disk induced in the realignment process, and the two curved radio jets \w2332\ are the relic of the early realignment stage.

The other plausible scenario involves two SMBHs residing in the disk-like host of \w2332 as a result of a major galaxy-galaxy merger. Based on numerical simulations, disk formation takes $\sim$ 1--2 Gyr after close encounter of the major merging event of the parent gas-rich galaxy pair \citep{2006ApJ...645..986R}. This period of time is enough for two SMBHs to sink down to the center of the merged system \citep{1980Natur.287..307B}, and reach the stalling separation at a distance of order pc or less. In this case, double-peaked Balmer lines arise from a long-lived eccentricity or single-arm spiral disk induced by the dynamical interaction with the second SMBH within a misaligned orbital plane, as discussed by \cite{1995ApJ...438..610E,2008ApJS..174..455B}. The binary interaction could be responsible for disk warping or a precession of the plane of the disk, as indicated by the mismatched inclination angles of the disk between now (implied by the spectroscopic models) and a few hundred thousand years ago (implied by the close-to-equal brightness of the primary jets on the two sides), and leaves the curved radio jets \citep[i.e. jet precession;][]{1988ApJ...334...95R} as well as the enigmatic radio features.

This scenario might apply not only in \w2332, but also in some DPEs. 
A recent VLBI survey of SDSS DPEs with FIRST detections shows that 2 out of 6 studied DPEs have compact radio cores at kilo-parsec scale separations (R. Deane, in prep.). The radio galaxy 4C37.11 contains an SMBH binary system with 7.3 pc separation and $\sim 1.5 \times 10^8\;M_{\sun}$ in mass \citep{2006ApJ...646...49R}. In addition, there are several radio-bright DPEs in which double nuclei with separations of a few kpc are seen in high resolution radio mapping, such as SDSS J1536+0441 \citep{2009ApJ...699L..22W,2010ApJ...714L.271B}, SDSS J1502+1115 \citep{2011ApJ...740L..44F}, and SDSS J1425+3231 \citep{2012MNRAS.425.1185F}. These results indicate that some DPE systems indeed host dual AGNs.

The current resolution limits of radio and millimeter interferometers, such as VLBI networks and ALMA, reach $\sim$ 5 mas \citep{2008RPPh...71f6901M,2009IEEEP..97.1463W}. At this resolution, compact radio/mm sources with projected distances $>$ 25~pc at $z = 0.3447$ can be resolved. Although this high contrast aperture synthesis imaging can recover high surface brightness structures such as highly collimated jets near their launch point, 25~pc physical resolution can not resolve an SMBH binary at ``intermediate'' merging stage with the stalling separations of 0.01--1~pc \citep{1980Natur.287..307B,2002MNRAS.331..935Y}. Current resolutions can, however, study an early stage SMBH merger, when the scattering of circumnuclear stars is the dominant process for losing angular momentum.

\section{Summary}\label{sec:summary}

We report the discovery of \w2332, a radio galaxy at $z = 0.3447$ with extraordinary radio morphology and optical spectroscopic features. We summarize the observational facts as follows:\\
(1) Exhibiting hybrid features between FR-I and FR-II morphology, the radio continuum map shows a winding Doppler-boosted FR-II jet, a pair of curved jets, a bright core, and complex emission within 25 kpc. \\
(2) The host galaxy of \w2332, unlike the elliptical hosts of most FR-II objects, is disk-like.\\
(3) We see long-term 3--20 yr optical-near-IR brightness variations from the AGN component.\\
(4) The optical spectrum of \w2332 shows offsets of $\sim$ 3800 km\,s$^{-1}$ between the centroids of the broad and narrow line components of \Ha\ and \Hb.

We consider the following two scenarios to explain these unusual properties: \\
(1) \w2332 has a single spinning SMBH misaligned with its accretion disk. The realignment of the accretion disk due to dynamical drag produces the broad-band Balmer profiles, leaving reorientated jets in the process, and making the complicated radio morphology. The cause of the initial misalignment is unknown. \\
(2) \w2332 has a SMBH binary with a separation on the order of a pc. The dynamical interaction between two SMBHs induces the warping or spiral pattern in the disk of one SMBH which generates the broad line profile. The curved jets and radio complex within the central 25 kpc reflect some perturbation from the second SMBH on the accretion disk.

Our results and analysis can neither rule out nor confirm either of these two scenarios. The real nature of \w2332 remains a mystery. Further investigations will be required to resolve the complicated structure within 5\arcsec\ radius of the core and confirm their nature, and even probe the sub-components in the core if there is any. We are currently pursuing the high-resolution radio mapping of \w2332 to investigate our hypotheses of the radio morphology formation, and also monitor its optical spectrum and photometry. If our hypothesis is confirmed, \w2332 will be an exciting system to study the physics of an SMBH merger event.

\acknowledgments
The authors thank the anonymous referee for the constructive comments and inspiring suggestions through out the whole paper. We acknowledge Joaquin Vieira for verifying the \w2332\ non-detection in the SPT survey map. We thank Roger Deane for sharing his VLBI work prior to the publication. We also appreciate the comments and suggestions by Colin Lonsdale, and the discussions with Michael Eracleous and Laura Blecha in ``Binary Black Holes and Dual AGN'' Meeting in Tucson 2012. RJA is supported by an appointment to the NASA Postdoctoral Program at the Jet Propulsion Laboratory, administered by Oak Ridge Associated Universities through a contract with NASA. This publication makes use of data products from the {\it Wide-field Infrared Survey Explorer}, which is a joint project of the University of California, Los Angeles, and the Jet Propulsion Laboratory, California Institute of Technology, funded by the National Aeronautics and Space Administration. Based on observations obtained at the Gemini Observatory, which is operated by the Association of Universities for Research in Astronomy, Inc., under a cooperative agreement with the NSF on behalf of the Gemini partnership: the National Science Foundation (United States), the Science and Technology Facilities Council (United Kingdom), the National Research Council (Canada), CONICYT (Chile), the Australian Research Council (Australia), Ministrio da Cincia, Tecnologia e Inovao (Brazil) and Ministerio de Ciencia, Tecnologa e Innovacin Productiva (Argentina).The Australia Telescope is funded by the Commonwealth of Australia for operation as a National Facility managed by CSIRO. This research has made use of the NASA/IPAC Extragalactic Database (NED) which is operated by the Jet Propulsion Laboratory, California Institute of Technology, under contract with the National Aeronautics and Space Administration. This research has made use of the NASA/ IPAC Infrared Science Archive, which is operated by the Jet Propulsion Laboratory, California Institute of Technology, under contract with the National Aeronautics and Space Administration. 

{\it Facilities:} \facility{ATCA (CABB)}, \facility{{\it Wide-field Infrared Survey Explorer}}, \facility{{\it Spitzer Space Telescope} (IRAC)}, \facility{SOAR (SOI imager)}, \facility{Gemini:South (GMOS spectrograph)}, \facility{{\it CXO} (ACIS-I)}.\\

\def\nar{NewAR}
\def\arxiv{arXiv}


\begin{thebibliography}{}
\bibitem[Abdo et al.(2010)]{2010ApJ...716...30A} Abdo, A.~A., Ackermann, M., Agudo, I., et al.\ 2010, \apj, 716, 30 
\bibitem[Alexander \& Hickox(2012)]{2012NewAR..56...93A} Alexander, D.~M., \& Hickox, R.~C.\ 2012, New Astronomy, 56, 93 
\bibitem[Anantharamaiah et al.(2000)]{2000ApJ...537..613A} Anantharamaiah, K.~R., Viallefond, F., Mohan, N.~R., Goss, W.~M., \& Zhao, J.~H.\ 2000, \apj, 537, 613 
\bibitem[Andersson et al.(2011)]{2011ApJ...738...48A} Andersson, K., Benson, B.~A., Ade, P.~A.~R., et al.\ 2011, \apj, 738, 48 
\bibitem[Assef et al.(2010)]{2010ApJ...713..970A} Assef, R.~J., Kochanek, C.~S., Brodwin, M., et al.\ 2010, \apj, 713, 970 
\bibitem[Bardeen \& Petterson(1975)]{1975ApJ...195L..65B} Bardeen, J.~M., \& Petterson, J.~A.\ 1975, \apjl, 195, L65 
\bibitem[Barrows et al.(2011)]{2011NewA...16..122B} Barrows, R.~S., Lacy, C.~H.~S., Kennefick, D., Kennefick, J., \& Seigar, M.~S.\ 2011, New Astronomy, 16, 122
\bibitem[Barrows et al.(2012)]{2012ApJ...744....7B} Barrows, R.~S., Stern, D., Madsen, K., et al.\ 2012, \apj, 744, 7 
\bibitem[Baumgarte \& Shapiro(2011)]{2011PhT....64j..32B} Baumgarte, T.~W., \& Shapiro, S.~L.\ 2011, Physics Today, 64, 100000 
\bibitem[Begelman et al.(1980)]{1980Natur.287..307B} Begelman, M.~C., Blandford, R.~D., \& Rees, M.~J.\ 1980, \nat, 287, 307 
\bibitem[Bohlin et al.(1978)]{1978ApJ...224..132B} Bohlin, R.~C., Savage, B.~D., \& Drake, J.~F.\ 1978, \apj, 224, 132 
\bibitem[Bogdanovi{\'c} et al.(2008)]{2008ApJS..174..455B} Bogdanovi{\'c}, T., Smith, B.~D., Sigurdsson, S., \& Eracleous, M.\ 2008, \apjs, 174, 455 
\bibitem[Bondi \& P{\'e}rez-Torres(2010)]{2010ApJ...714L.271B} Bondi, M., \& P{\'e}rez-Torres, M.-A.\ 2010, \apjl, 714, L271 
\bibitem[Bonning et al.(2007)]{2007ApJ...666L..13B} Bonning, E.~W., Shields, G.~A., \& Salviander, S.\ 2007, \apjl, 666, L13 
\bibitem[Boroson \& Lauer(2009)]{2009Natur.458...53B} Boroson, T.~A., \& Lauer, T.~R.\ 2009, \nat, 458, 53 
\bibitem[Campanelli et al.(2007)]{2007ApJ...659L...5C} Campanelli, M., Lousto, C., Zlochower, Y., \& Merritt, D.\ 2007, \apjl, 659, L5 
\bibitem[Cannizzo et al.(1990)]{1990ApJ...351...38C} Cannizzo, J.~K., Lee, H.~M., \& Goodman, J.\ 1990, \apj, 351, 38 
\bibitem[Capetti et al.(2002)]{2002A&A...394...39C} Capetti, A., Zamfir, S., Rossi, P., et al.\ 2002, \aap, 394, 39 
\bibitem[Chen \& Halpern(1989)]{1989ApJ...344..115C} Chen, K., \& Halpern, J.~P.\ 1989, \apj, 344, 115 
\bibitem[Cheung et al.(2009)]{2009ApJS..181..548C} Cheung, C.~C., Healey, S.~E., Landt, H., Verdoes Kleijn, G., \& Jord{\'a}n, A.\ 2009, \apjs, 181, 548 
\bibitem[Civano et al.(2010)]{2010ApJ...717..209C} Civano, F., Elvis, M., Lanzuisi, G., et al.\ 2010, \apj, 717, 209 
\bibitem[Collin-Souffrin \& Dumont(1989)]{1989A&A...213...29C} Collin-Souffrin, S., \& Dumont, A.~M.\ 1989, \aap, 213, 29 
\bibitem[Comerford et al.(2012)]{2012ApJ...753...42C} Comerford, J.~M., Gerke, B.~F., Stern, D., et al.\ 2012, \apj, 753, 42 
\bibitem[Cutri et al.(2012)]{2012wise.rept....1C} Cutri, R.~M., Wright, E.~L., Conrow, T., et al.\ 2012, Explanatory Supplement to the WISE All-Sky Data Release Products, 1 
\bibitem[Dennett-Thorpe et al.(2002)]{2002MNRAS.330..609D} Dennett-Thorpe, J., Scheuer, P.~A.~G., Laing, R.~A., et al.\ 2002, \mnras, 330, 609 
\bibitem[Dotti et al.(2009)]{2009MNRAS.396.1640D} Dotti, M., Ruszkowski, M., Paredi, L., et al.\ 2009, \mnras, 396, 1640 
\bibitem[Dotti et al.(2007)]{2007MNRAS.379..956D} Dotti, M., Colpi, M., Haardt, F., \& Mayer, L.\ 2007, \mnras, 379, 956 
\bibitem[Dumont \& Collin-Souffrin(1990a)]{1990A&A...229..313D} Dumont, A.~M., \& Collin-Souffrin, S.\ 1990, \aap, 229, 313 \bibitem[Dressler et al.(2011)]{2011PASP..123..288D} Dressler, A., Bigelow, B., Hare, T., et al.\ 2011, \pasp, 123, 288 
\bibitem[Dumont \& Collin-Souffrin(1990b)]{1990A&AS...83...71D} Dumont, A.~M., \& Collin-Souffrin, S.\ 1990, \aaps, 83, 71 
\bibitem[Ekers et al.(1978)]{1978A&A....69L..21E} Ekers, R.~D., Goss, W.~M., Kotanyi, C.~G., \& Skellern, D.~J.\ 1978, \aap, 69, L21 
\bibitem[Emonts et al.(2008)]{2008MNRAS.387..197E} Emonts, B.~H.~C., Morganti, R., Oosterloo, T.~A., et al.\ 2008, \mnras, 387, 197 
\bibitem[Emonts et al.(2009)]{2009MNRAS.396.1522E} Emonts, B.~H.~C., Tadhunter, C.~N., Morganti, R., et al.\ 2009, \mnras, 396, 1522 
\bibitem[Eracleous et al.(2012)]{2012ApJS..201...23E} Eracleous, M., Boroson, T.~A., Halpern, J.~P., \& Liu, J.\ 2012, \apjs, 201, 23 
\bibitem[Eracleous et al.(1995)]{1995ApJ...438..610E} Eracleous, M., Livio, M., Halpern, J.~P., \& Storchi-Bergmann, T.\ 1995, \apj, 438, 610 
\bibitem[Eracleous \& Halpern(2003)]{2003ApJ...599..886E} Eracleous, M., \& Halpern, J.~P.\ 2003, \apj, 599, 886 
\bibitem[Eracleous et al.(2009)]{2009NewAR..53..133E} Eracleous, M., Lewis, K.~T., \& Flohic, H.~M.~L.~G.\ 2009, \nar, 53, 133 
\bibitem[Escala et al.(2004)]{2004ApJ...607..765E} Escala, A., Larson, R.~B., Coppi, P.~S., \& Mardones, D.\ 2004, \apj, 607, 765 
\bibitem[Escala et al.(2005)]{2005ApJ...630..152E} Escala, A., Larson, R.~B., Coppi, P.~S., \& Mardones, D.\ 2005, \apj, 630, 152 
\bibitem[Fanaroff \& Riley(1974)]{1974MNRAS.167P..31F} Fanaroff, B.~L., \& Riley, J.~M.\ 1974, \mnras, 167, 31P 
\bibitem[Frey et al.(2012)]{2012MNRAS.425.1185F} Frey, S., Paragi, Z., An, T., \& Gab{\'a}nyi, K.~{\'E}.\ 2012, \mnras, 425, 1185 
\bibitem[Fu et al.(2011a)]{2011ApJ...733..103F} Fu, H., Myers, A.~D., Djorgovski, S.~G., \& Yan, L.\ 2011, \apj, 733, 103 
\bibitem[Fu et al.(2011b)]{2011ApJ...740L..44F} Fu, H., Zhang, Z.-Y., Assef, R.~J., et al.\ 2011, \apjl, 740, L44
\bibitem[Fu et al.(2012)]{2012ApJ...745...67F} Fu, H., Yan, L., Myers, A.~D., et al.\ 2012, \apj, 745, 67 
\bibitem[Gezari et al.(2007)]{2007ApJS..169..167G} Gezari, S., Halpern, J.~P., \& Eracleous, M.\ 2007, \apjs, 169, 167 
\bibitem[Gonz{\'a}lez et al.(2007)]{2007PhRvL..98w1101G} Gonz{\'a}lez, J.~A., Hannam, M., Sperhake, U., Br{\"u}gmann, B., \& Husa, S.\ 2007, Physical Review Letters, 98, 231101 
\bibitem[Gopal-Krishna \& Wiita(2000)]{2000A&A...363..507G} Gopal-Krishna, \& Wiita, P.~J.\ 2000, \aap, 363, 507 
\bibitem[Gopal-Krishna et al.(2010)]{2010arXiv1008.0789G} Gopal-Krishna, Biermann, P.~L., Gergely, L.~{\'A}., \& Wiita, P.~J.\ 2010, \arxiv:1008.0789 
\bibitem[Gopal-Krishna et al.(2003)]{2003ApJ...594L.103G} Gopal-Krishna, Biermann, P.~L., \& Wiita, P.~J.\ 2003, \apjl, 594, L103 
\bibitem[Green et al.(2010)]{2010ApJ...710.1578G} Green, P.~J., Myers, A.~D., Barkhouse, W.~A., et al.\ 2010, \apj, 710, 1578 
\bibitem[Gregory et al.(1994)]{1994ApJS...90..173G} Gregory, P.~C., Vavasour, J.~D., Scott, W.~K., \& Condon, J.~J.\ 1994, \apjs, 90, 173 
\bibitem[G{\"u}ltekin et al.(2009)]{2009ApJ...706..404G} G{\"u}ltekin, K., Cackett, E.~M., Miller, J.~M., et al.\ 2009, \apj, 706, 404 
\bibitem[Hambly et al.(2001)]{2001MNRAS.326.1295H} Hambly, N.~C., Irwin, M.~J., \& MacGillivray, H.~T.\ 2001, \mnras, 326, 1295 
\bibitem[Healy et al.(2009)]{2009PhRvL.102d1101H} Healy, J., Herrmann, F., Hinder, I., et al.\ 2009, Physical Review Letters, 102, 041101 
\bibitem[Heckman et al.(1982)]{1982ApJ...262..529H} Heckman, T.~M., Miley, G.~K., Balick, B., van Breugel, W.~J.~M., \& Butcher, H.~R.\ 1982, \apj, 262, 529
\bibitem[High et al.(2010)]{2010ApJ...723.1736H} High, F.~W., Stalder, B., Song, J., et al.\ 2010, \apj, 723, 1736 
\bibitem[Hodges-Kluck et al.(2010)]{2010ApJ...710.1205H} Hodges-Kluck, E.~J., Reynolds, C.~S., Cheung, C.~C., \& Miller, M.~C.\ 2010, \apj, 710, 1205 
\bibitem[Hook et al.(2004)]{2004PASP..116..425H} Hook, I.~M., J{\o}rgensen, I., Allington-Smith, J.~R., et al.\ 2004, \pasp, 116, 425 
\bibitem[Hota et al.(2011)]{2011MNRAS.417L..36H} Hota, A., Sirothia, S.~K., Ohyama, Y., et al.\ 2011, \mnras, 417, L36 
\bibitem[Iguchi et al.(2010)]{2010ApJ...724L.166I} Iguchi, S., Okuda, T., \& Sudou, H.\ 2010, \apjl, 724, L166 
\bibitem[Jarrett et al.(2011)]{2011ApJ...735..112J} Jarrett, T.~H., Cohen, M., Masci, F., et al.\ 2011, \apj, 735, 112 
\bibitem[Jenet et al.(2004)]{2004ApJ...606..799J} Jenet, F.~A., Lommen, A., Larson, S.~L., \& Wen, L.\ 2004, \apj, 606, 799 
\bibitem[Ju et al.(2013)]{2013arXiv.1306.4987J} Ju, W., Greene, J.~E., Rafikov, R.~R., Bickerton, S.~J., \& Badenes, C.\ 2013, \arxiv:1306.4987 
\bibitem[Keel et al.(2006)]{2006AJ....132.2233K} Keel, W.~C., White, R.~E., III, Owen, F.~N., \& Ledlow, M.~J.\ 2006, \aj, 132, 2233 
\bibitem[Kellermann et al.(1989)]{1989AJ.....98.1195K} Kellermann, K.~I., Sramek, R., Schmidt, M., Shaffer, D.~B., \& Green, R.\ 1989, \aj, 98, 1195 
\bibitem[Kennicutt(1998)]{1998ARA&A..36..189K} Kennicutt, R.~C., Jr.\ 1998, \araa, 36, 189 
\bibitem[Komossa et al.(2003)]{2003ApJ...582L..15K} Komossa, S., Burwitz, V., Hasinger, G., et al.\ 2003, \apjl, 582, L15 
\bibitem[Komossa(2012)]{2012AdAst2012E..14K} Komossa, S.\ 2012, Advances in Astronomy, 2012, id. 364973
\bibitem[Landt et al.(2010)]{2010MNRAS.408.1103L} Landt, H., Cheung, C.~C., \& Healey, S.~E.\ 2010, \mnras, 408, 1103 
\bibitem[Leahy \& Parma(1992)]{1992ersf.meet..307L} Leahy, J.~P., \& Parma, P.\ 1992, Extragalactic Radio Sources.~From Beams to Jets, 307 
\bibitem[Leahy \& Williams(1984)]{1984MNRAS.210..929L} Leahy, J.~P., \& Williams, A.~G.\ 1984, \mnras, 210, 929 
\bibitem[Ledlow et al.(2001)]{2001ApJ...552..120L} Ledlow, M.~J., Owen, F.~N., Yun, M.~S., \& Hill, J.~M.\ 2001, \apj, 552, 120 
\bibitem[Lewis et al.(2010)]{2010ApJS..187..416L} Lewis, K.~T., Eracleous, M., \& Storchi-Bergmann, T.\ 2010, \apjs, 187, 416 
\bibitem[Maloney et al.(1996)]{1996ApJ...472..582M} Maloney, P.~R., Begelman, M.~C., \& Pringle, J.~E.\ 1996, \apj, 472, 582 
\bibitem[Mao et al.(2010)]{2010IAUS..267..119M} Mao, M.~Y., Norris, R.~P., Sharp, R., \& Lovell, J.~E.~J.\ 2010, IAU Symposium, 267, 119 
\bibitem[Matthews et al.(1964)]{1964ApJ...140...35M} Matthews, T.~A., Morgan, W.~W., \& Schmidt, M.\ 1964, \apj, 140, 35 
\bibitem[Massaro et al.(2012)]{2012ApJ...750..138M} Massaro, F., D'Abrusco, R., Tosti, G., et al.\ 2012, \apj, 750, 138 
\bibitem[Mauch et al.(2003)]{2003MNRAS.342.1117M} Mauch, T., Murphy, T., Buttery, H.~J., et al.\ 2003, \mnras, 342, 1117 
\bibitem[McConnell et al.(2011)]{2011Natur.480..215M} McConnell, N.~J., Ma, C.-P., Gebhardt, K., et al.\ 2011, \nat, 480, 215 
\bibitem[McConnell et al.(2012)]{2012ApJ...756..179M} McConnell, N.~J., Ma, C.-P., Murphy, J.~D., et al.\ 2012, \apj, 756, 179 
\bibitem[Meier et al.(2001)]{2001Sci...291...84M} Meier, D.~L., Koide, S., \& Uchida, Y.\ 2001, Science, 291, 84 
\bibitem[Merloni et al.(2003)]{2003MNRAS.345.1057M} Merloni, A., Heinz, S., \& di Matteo, T.\ 2003, \mnras, 345, 1057 
\bibitem[Merritt \& Ekers(2002)]{2002Sci...297.1310M} Merritt, D., \& Ekers, R.~D.\ 2002, Science, 297, 1310 
\bibitem[Merritt \& Milosavljevi{\'c}(2005)]{2005LRR.....8....8M} Merritt, D., \& Milosavljevi{\'c}, M.\ 2005, Living Reviews in Relativity, 8, 8 
\bibitem[Middelberg \& Bach(2008)]{2008RPPh...71f6901M} Middelberg, E., \& Bach, U.\ 2008, Reports on Progress in Physics, 71, 066901 
\bibitem[Middelberg et al.(2008)]{2008AJ....135.1276M} Middelberg, E., Norris, R.~P., Cornwell, T.~J., et al.\ 2008, \aj, 135, 1276 
\bibitem[Morganti et al.(2011)]{2011A&A...535A..97M} Morganti, R., Holt, J., Tadhunter, C., et al.\ 2011, \aap, 535, A97 
\bibitem[Norris et al.(2006)]{2006AJ....132.2409N} Norris, R.~P., Afonso, J., Appleton, P.~N., et al.\ 2006, \aj, 132, 2409 
\bibitem[Norris et al.(2012)]{2012MNRAS.422.1453N} Norris, R.~P., Lenc, E., Roy, A.~L., \& Spoon, H.\ 2012, \mnras, 422, 1453 
\bibitem[Ogilvie(2001)]{2001MNRAS.325..231O} Ogilvie, G.~I.\ 2001, \mnras, 325, 231 
\bibitem[Owen et al.(1985)]{1985ApJ...294L..85O} Owen, F.~N., O'Dea, C.~P., Inoue, M., \& Eilek, J.~A.\ 1985, \apjl, 294, L85 
\bibitem[Owen \& Laing(1989)]{1989MNRAS.238..357O} Owen, F.~N., \& Laing, R.~A.\ 1989, \mnras, 238, 357 
\bibitem[Owen \& White(1991)]{1991MNRAS.249..164O} Owen, F.~N., \& White, R.~A.\ 1991, \mnras, 249, 164 
\bibitem[Peng et al.(2002)]{2002AJ....124..266P} Peng, C.~Y., Ho, L.~C., Impey, C.~D., \& Rix, H.-W.\ 2002, \aj, 124, 266 
\bibitem[Peng et al.(2010)]{2010AJ....139.2097P} Peng, C.~Y., Ho, L.~C., Impey, C.~D., \& Rix, H.-W.\ 2010, \aj, 139, 2097 
\bibitem[Popovi{\'c}(2012)]{2012NewAR..56...74P} Popovi{\'c}, L.~{\v C}.\ 2012, \nar, 56, 74 
\bibitem[Rachford et al.(2002)]{2002ApJ...577..221R} Rachford, B.~L., Snow, T.~P., Tumlinson, J., et al.\ 2002, \apj, 577, 221 
\bibitem[Rees(1988)]{1988Natur.333..523R} Rees, M.~J.\ 1988, \nat, 333, 523 
\bibitem[Robertson et al.(2006)]{2006ApJ...645..986R} Robertson, B., Bullock, J.~S., Cox, T.~J., et al.\ 2006, \apj, 645, 986 
\bibitem[Rodriguez et al.(2006)]{2006ApJ...646...49R} Rodriguez, C., Taylor, G.~B., Zavala, R.~T., et al.\ 2006, \apj, 646, 49 
\bibitem[Roos(1988)]{1988ApJ...334...95R} Roos, N.\ 1988, \apj, 334, 95 
\bibitem[Saripalli \& Subrahmanyan(2009)]{2009ApJ...695..156S} Saripalli, L., \& Subrahmanyan, R.\ 2009, \apj, 695, 156 
\bibitem[Shen et al.(2013)]{2013arXiv1306.4330S} Shen, Y., Liu, X., Loeb, A., \& Tremaine, S.\ 2013, \arxiv:1306.4330
\bibitem[Shields et al.(2009)]{2009ApJ...696.1367S} Shields, G.~A., Bonning, E.~W., \& Salviander, S.\ 2009, \apj, 696, 1367 
\bibitem[Shields et al.(2009)]{2009ApJ...707..936S} Shields, G.~A., Rosario, D.~J., Smith, K.~L., et al.\ 2009, \apj, 707, 936 
\bibitem[Skrutskie et al.(2006)]{2006AJ....131.1163S} Skrutskie, M.~F., Cutri, R.~M., Stiening, R., et al.\ 2006, \aj, 131, 1163 
\bibitem[Steinhardt et al.(2012)]{2012arXiv1209.1635S} Steinhardt, C.~L., Schramm, M., Silverman, J.~D., et al.\ 2012, \arxiv:1209.1635 
\bibitem[Stern et al.(2002)]{2002AJ....123.2223S} Stern, D., Tozzi, P., Stanford, S.~A., et al.\ 2002, \aj, 123, 2223 
\bibitem[Stern et al.(2012)]{2012ApJ...753...30S} Stern, D., Assef, R.~J., Benford, D.~J., et al.\ 2012, \apj, 753, 30 
\bibitem[Storchi-Bergmann et al.(2003)]{2003ApJ...598..956S} Storchi-Bergmann, T., Nemmen da Silva, R., Eracleous, M., et al.\ 2003, \apj, 598, 956 
\bibitem[Sudou et al.(2003)]{2003Sci...300.1263S} Sudou, H., Iguchi, S., Murata, Y., \& Taniguchi, Y.\ 2003, Science, 300, 1263 
\bibitem[Syer \& Clarke(1992)]{1992MNRAS.255...92S} Syer, D., \& Clarke, C.~J.\ 1992, \mnras, 255, 92 
\bibitem[Tang \& Grindlay(2009)]{2009ApJ...704.1189T} Tang, S., \& Grindlay, J.\ 2009, \apj, 704, 1189 
\bibitem[Tsalmantza et al.(2011)]{2011ApJ...738...20T} Tsalmantza, P., Decarli, R., Dotti, M., \& Hogg, D.~W.\ 2011, \apj, 738, 20 
\bibitem[Ulrich et al.(1997)]{1997ARA&A..35..445U} Ulrich, M.-H., Maraschi, L., \& Urry, C.~M.\ 1997, \araa, 35, 445 
\bibitem[van Breugel et al.(1984)]{1984ApJ...277...82V} van Breugel, W., Heckman, T., Butcher, H., \& Miley, G.\ 1984, \apj, 277, 82 
\bibitem[Vanderlinde et al.(2010)]{2010ApJ...722.1180V} Vanderlinde, K., Crawford, T.~M., de Haan, T., et al.\ 2010, \apj, 722, 1180 
\bibitem[Walker et al.(2003)]{2003SPIE.4841..286W} Walker, A.~R., Boccas, M., Bonati, M., et al.\ 2003, \procspie, 4841, 286 
\bibitem[Wootten \& Thompson(2009)]{2009IEEEP..97.1463W} Wootten, A., \& Thompson, A.~R.\ 2009, IEEE Proceedings, 97, 1463 
\bibitem[Worrall et al.(1995)]{1995ApJ...449...93W} Worrall, D.~M., Birkinshaw, M., \& Cameron, R.~A.\ 1995, \apj, 449, 93 
\bibitem[Wright et al.(1994)]{1994ApJS...91..111W} Wright, A.~E., Griffith, M.~R., Burke, B.~F., \& Ekers, R.~D.\ 1994, \apjs, 91, 111 
\bibitem[Wright et al.(2010)]{2010AJ....140.1868W} Wright, E.~L., Eisenhardt, P.~R.~M., Mainzer, A.~K., et al.\ 2010, \aj, 140, 1868 
\bibitem[Wrobel \& Laor(2009)]{2009ApJ...699L..22W} Wrobel, J.~M., \& Laor, A.\ 2009, \apjl, 699, L22 
\bibitem[Wu et al.(2008)]{2008MNRAS.389..213W} Wu, S.-M., Wang, T.-G., \& Dong, X.-B.\ 2008, \mnras, 389, 213 
\bibitem[Yokosawa \& Inoue(1985)]{1985PASJ...37..655Y} Yokosawa, M., \& Inoue, M.\ 1985, \pasj, 37, 655 
\bibitem[Yu(2002)]{2002MNRAS.331..935Y} Yu, Q.\ 2002, \mnras, 331, 935 
\end{thebibliography}
\end{document}